%% file: n.tex
\documentclass[]{elsart}

 \usepackage{graphicx}
 \usepackage{epsfig}

\usepackage{amssymb}
\usepackage{amsmath}
\usepackage{letterspace}
\usepackage{longtable}
\usepackage{textcomp}
\usepackage{array}

\begin{document}
\input{macros.tex}

\begin{frontmatter}

\input{n0.tex}
\end{frontmatter}


\setlongtables
\input{n1.tex}
\input{n2.tex}

\input{n3.tex}

\input{n4.tex}

\input{n5.tex}
\input{n6.tex}

\input{n7.tex}

\label{}

\bibliography{t}
\bibliographystyle{unsrt.bst}
\end{document}

%% file: macros.tex
\newcommand{\una}[2]{$#1\ \text{#2}$}
\newcommand{\unb}[2]{$#1\text{-#2}$}
\newcommand{\slsh}[1]{/\letterspace to -1.0\naturalwidth{#1}\ \ }
\newcommand{\bi}{\begin{itemize}}
\newcommand{\ei}{\end{itemize}}
\newcommand{\be}{\begin{enumerate}}
\newcommand{\ee}{\end{enumerate}}
\newcommand{\bd}{\begin{description}}
\newcommand{\ed}{\end{description}}
\def\alr{A_{LR}} 
\def\alrm{A_{LR}^{M\o ller}} 
\def\alrep{A_{LR}^{ep}} 
\def\alree{A_{LR}^{ee}} 
\def\alrb{^{\text{beam}}\text{A}_{\text{LR}}} 
\def\alrbt{^{beam}\!\widetilde{A}_{LR}} 
\def\hcs{helicity correlations}
\def\gp{F_P} 
\def\ga{F_A} 
\def\gv{F_V} 
\def\gm{F_M} 
\def\Gp{G_P}  
\def\Gv{G_V} 
\def\Ga{G_A} 
\def\vsig{\vec \sigma} 
\def\zhat{\hat z} 
\def\khat{\hat k} 
\def\cth{\khat \cdot \zhat} 
\def\gmu{\gamma_{\mu}} 
\def\g5{\gamma_5} 
\def\mmu{m_\mu} 
\def\Enu{E_{\nu}} 

\def\z0{Z^0}
\def\u{$u$}
\def\d{$d$}
\def\s{$s$}
\def\gu{$G_u$}
\def\gd{$G_d$}
\def\gs{$G_s$}
\def\gc{$G_c$}
\def\gA{$G_A$}
\def\gEp{G_E^p}
\def\gEn{G_E^n}
\def\gMp{G_M^p}
\def\gMn{G_M^n}
\def\gEpg{G_E^{p\gamma}}
\def\gEng{G_E^{n\gamma}}
\def\gMpg{G_M^{p\gamma}}
\def\gMng{G_M^{n\gamma}}
\def\gEMpg{G_{E,M}^{p\gamma}}
\def\gEMng{G_{E,M}^{n\gamma}}
\def\gEMpng{G_{E,M}^{pn\gamma}}
\def\gEMpz{G_{E,M}^{pZ}}
\def\gEMnz{G_{E,M}^{nZ}}
\def\gEMp{G_{E,M}^{p}}

\def\gEu{G_E^u}
\def\gMu{G_M^u}
\def\gEd{G_E^d}
\def\gMd{G_M^d}
\def\gEs{G_E^s}
\def\gMs{G_M^s}
\def\gAs{G_A^s}
\def\gEMu{G_{E,M}^u}
\def\gEMd{G_{E,M}^d}
\def\gEMs{G_{E,M}^s}
\def\gAz{G_A^Z}
\def\gPz{G_P^Z}

\def\e158{$\text{E-158}$}
\def\t437{$\text{T-437}$}
\def\hap{HAPPEX}
\def\lra{\leftrightarrow}
\def\mo{M\o ller}
\def\sw{\sin^2 \theta_W}
\def\q2{Q^2}

\def\NCA{{\em Nuovo Cimento} A}
\def\PHYS{{ Physica}}
\def\NPA{{ Nucl. Phys.} A}
\def\MATH{{ J. Math. Phys.}}
\def\PRO{{Prog. Theor. Phys.}}
\def\NPB{{ Nucl. Phys.} B}
\def\PLA{{ Phys. Lett.} A}
\def\PLB{{ Phys. Lett.} B}
\def\PLD{{ Phys. Lett.} D}
\def\PL{{ Phys. Lett.}}
\def\PRL{Phys. Rev. Lett.}
\def\PREV{ Phys. Rev.}
\def\PREP{ Phys. Rep.}
\def\PRA{{ Phys. Rev.} A}
\def\PRD{{ Phys. Rev.} D}
\def\PRC{{ Phys. Rev.} C}
\def\PRB{{ Phys. Rev.} B}
\def\ZPC{{ Z. Phys.} C}
\def\ZPA{{ Z. Phys.} A}
\def\ANNP{ Ann. Phys. (N.Y.)}
\def\RMP{{ Rev. Mod. Phys.}}
\def\CHEM{{ J. Chem. Phys.}}
\def\INT{{ Int. J. Mod. Phys.} E}
\def\NIM{{ Nucl. Instr. Methods A}}

%% file: n0.tex


\title{SLAC's Polarized Electron Source Laser System and Minimization of Electron Beam Helicity Correlations for the $\text{E-158}$ Parity Violation Experiment}


\author[e]{T. B. Humensky,\corauthref{cor}}
\corauth[cor]{Corresponding Author.}
\ead{humensky@slac.stanford.edu}
\author[b]{R. Alley,}
\author[a]{A. Brachmann,}
\author[a]{M. J. Browne,}
\author[c]{G. D. Cates,}
\author[a]{J. Clendenin,}
\author[a]{J. deLamare,}
\author[a]{J. Frisch,}
\author[a]{T. Galetto,}
\author[d]{E. W. Hughes,}
\author[f]{K. S. Kumar,}
\author[d]{P. Mastromarino,}
\author[a]{J. Sodja,}
\author[h]{P. A. Souder,}
\author[a]{J. Turner,} and 
\author[a]{M. Woods}

\address[e]{Princeton University, Princeton, NJ}
\address[b]{Illumina, Inc., San Diego, CA}
\address[a]{Stanford Linear Accelerator Center, Menlo Park, CA}
\address[c]{University of Virginia, Charlottesville, VA}
\address[d]{California Institute of Technology, Pasadena, CA}
\address[f]{University of Massachusetts, Amherst, MA}
\address[h]{Syracuse University, Syracuse, NY}

\begin{abstract}

SLAC $\text{E-158}$ is an experiment designed to make the first measurement of parity violation in M\o ller scattering.  $\text{E-158}$ will measure the right-left cross-section asymmetry, $A_{LR}^{M\o ller}$, in the elastic scattering of a 45-GeV polarized electron beam off unpolarized electrons in a liquid hydrogen target. $\text{E-158}$ plans to measure the expected Standard Model asymmetry of $\sim 10^{-7}$ to an accuracy of better than 10$^{-8}$.  To make this measurement, the polarized electron source requires for operation an intense circularly polarized laser beam and the ability to quickly switch between right- and left-helicity polarization states with minimal right-left helicity-correlated asymmetries in the resulting beam parameters (intensity, position, angle, spot size, and energy), $\alrb$'s. This laser beam is produced by a unique SLAC-designed flashlamp-pumped Ti:Sapphire laser and is propagated through a carefully designed set of polarization optics.  We analyze the transport of nearly circularly polarized light through the optical system and identify several mechanisms that generate $\alrb$'s.  We show that the dominant effects depend linearly on particular polarization phase shifts in the optical system.  We present the laser system design and a discussion of the suppression and control of $\alrb$'s.  We also present results on beam performance from engineering and physics runs for $\text{E-158}$.

\end{abstract}

\begin{keyword}
helicity-correlated asymmetry \sep \mo\ scattering \sep parity violation \sep polarized electrons \sep Ti:Sapphire laser \sep Standard Model

\PACS 29.25.Bx \sep 29.27.Hj \sep 42.25.Ja
\end{keyword}

%% file: n1.tex
\setcounter{tocdepth}{4}
\tableofcontents
\section{$\mathbf{Introduction}$}
\renewcommand{\thefootnote}{\fnsymbol{footnote}}
\setcounter{footnote}{0}
The Stanford Linear Accelerator Center (SLAC) has a distinguished history of providing polarized electron beams for use in high energy physics experiments, including important tests of the Standard Model of particle physics and detailed studies of the spin structure of nucleons \cite{Prescott79,SLCrev01,spinrev99}. SLAC's polarized electron source is based on photoemission from a strained GaAs cathode pumped by an intense, circularly polarized laser beam \cite{Alley95,turner}.  Two laser systems exist to pump the cathode: a Nd:YLF-pumped Ti:Sapphire laser (the ``YLF:Ti'') that generates short ($2\text{-ns}$) pulses of electrons for SLAC's Positron Electron Project (PEP) rings, and a flashlamp-pumped Ti:Sapphire laser (the ``Flash:Ti'') used to generate $270\text{-ns}$ pulses for use in fixed target experiments at SLAC's ``End Station A'' such as E-158, described briefly below. 

E-158 will make the first measurement of parity violation in M\o ller scattering by measuring the asymmetry in the cross section for elastic scattering of longitudinally polarized electrons with an energy of 45 GeV off of an unpolarized electron target:
\begin{equation}  A_{LR}^{\mo} = \frac{\sigma_R - \sigma_L}{\sigma_R +
\sigma_L},
\label{eq:asy}
 \end{equation}
where $\sigma_R$ ($\sigma_L$) is the cross section for incident right- (left-) helicity electrons \cite{e158,cipanp}. The experiment will make a stringent test of the Standard Model of particle physics and will also be sensitive to new physics beyond the Standard Model~\cite{cm}. The asymmetry will be measured to an accuracy of better than 10$^{-8}$, with the expected Standard Model asymmetry being approximately 10$^{-7}$.  

One critical challenge for E-158 is the suppression of helicity-correlated asymmetries in the parameters of the electron beam when it is incident on the liquid hydrogen target.  Helicity-correlated asymmetries in the electron beam must be held to very small levels to prevent them from contributing false asymmetries to the measurement at a significant level.  For instance, because typical fixed-target scattering cross sections are proportional to $\sin^{-4}\theta$ and the detector accepts electrons with scattering angles of $4.5-7.2\ \text{mrad}$, the scattered flux reaching the detector is strongly dependent on the position and angle of the beam at the target.  If, for example, over the length of the experiment the average beam position on target for right- and left-helicity pulses is different, a false asymmetry will be measured that is proportional to the magnitude of that difference.  The \mo\ physics asymmetry in equation~\ref{eq:asy} can be expressed in terms of measured detector and beam quantities.  Consider first a single pair of pulses corresponding to right- and left-helicity beam incident on the \e158 target.  For the case of small right-left differences (and neglecting background contributions), the asymmetry for a single pair $A^P_{LR}$ can be written as 
\begin{align}  A^P_{LR} = P_B \cdot \alrm 
&= \frac{\Delta D}{2D} - \frac{\Delta I}{2I} + \alpha_E \frac{\Delta E}{2E} + \sum_i\alpha_i\Delta X_i \\
&= \frac{\Delta D}{2D} -\, \alrbt, \notag
 \end{align}
where $P_B$ is the beam polarization, $D \propto \sigma$ is the average detected scattered flux for right- and left-helicity pulses, $I$ is the beam intensity, $E$ is the beam energy, the $X_i$ run over position and angle in $x$ and $y$, and $\alpha_E$ and the $\alpha_i$ are correlation coefficients between energy, position, and angle and the detector signal. These coefficients are measured simultaneously with data-taking.  $\Delta$ refers to the right-left difference in each of the above parameters.  We use the symbol $\alrbt$ to refer to the contribution to the measured detector asymmetry arising from helicity-correlated beam asymmetries.  We frequently use the acronym ``\,$\alrb$'' to refer to a helicity-correlated asymmetry (or difference) in the intensity, energy, position, angle, or other parameters of the electron beam.  On occasion we use the terms ``helicity correlation'' or ``helicity-correlated asymmetry'' to also refer to these beam asymmetries.  \e158 will accumulate many right-left pairs ($\sim3 \cdot 10^8$) over the length of the physics run.  Taking into account the detector's nonlinearity (projected to be $0.5\ \%$) and estimates of the sensitivity of the scattering cross section to energy, position, and angle (determined by measuring $\alpha_E$ and the $\alpha_i$), we estimate that the intensity asymmetry $A_I$, the energy asymmetry $A_E$, and the position and angle differences $D_{X(Y)}$ and $D_{X'(Y')}$ must be held below the following limits:
\begin{xalignat}{2}
\label{eq:lim}
A_{I} &= \langle\frac{I_R - I_L}{I_R + I_L}\rangle <2 \cdot 10^{-7},& A_{E} &= \langle\frac{E_R - E_L}{E_R + E_L}\rangle <2 \cdot 10^{-8}, \\
D_{X} &= \langle x_R - x_L\rangle <10 \text{ nm,}& D_{X'} &= \langle x'_R - x'_L\rangle <0.4 \text{ nrad,} \notag
\end{xalignat}
where the angled brackets denote averaging over all pairs.  Achieving these limits will keep contributions to the systematic error on $\alrm$ at the level of $1\ \text{ppb}$ or less from each $\alrb$ and keep the cumulative systematic error contribution from all $\alrb$'s at the level of $3\ \text{ppb}$, a level comfortably below the projected statistical error of $8\ \text{ppb}$.  The derivation of these limits is discussed in more detail in \cite{e158,t437}.  A major focus of the work presented in this paper is the implementation of a number of methods for controlling $\alrb$'s at a level that will allow achieving the requirements of equations~\ref{eq:lim}.  

\begin{figure}
   \centering
   \begin{tabular}{c}
   \includegraphics{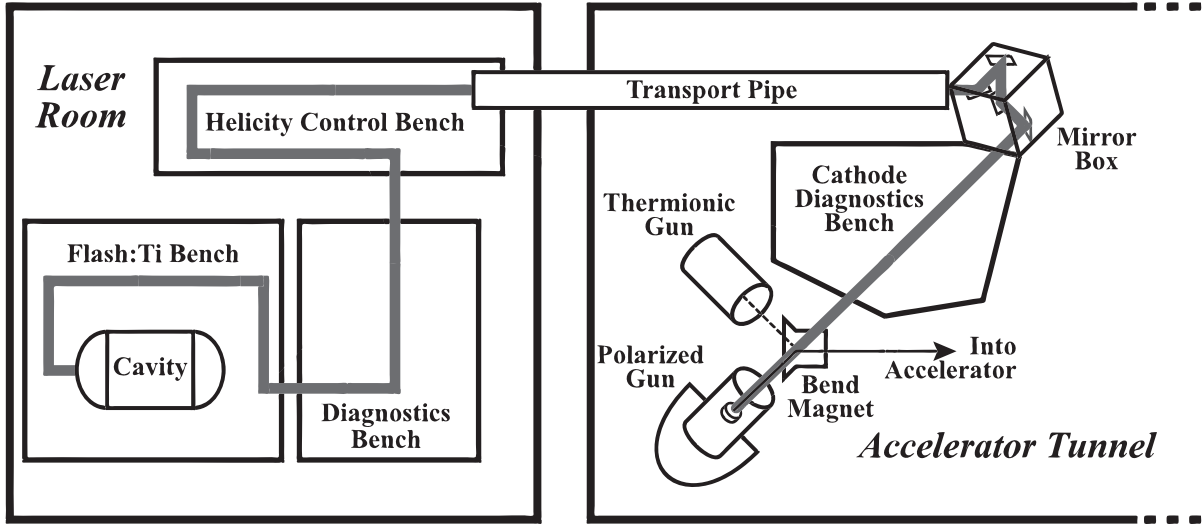}
   \end{tabular}
   \caption[example] 
   { \label{fig:overview} 
An overview of the Polarized Electron Source as it is configured for E-158.}
   \end{figure}

In this paper we describe the Flash:Ti laser and the accompanying optics system for the polarized electron source.  The Flash:Ti system was originally designed and commissioned in 1993 \cite{Alley95,Witte94} and has seen extensive upgrades in preparation for $\text{E-158}$.  Table~\ref{tab:flashti} summarizes the parameters of the Flash:Ti laser beam for \e158.  Likewise, the polarization and transport optics have also seen significant upgrades.  The focus of these upgrades has been to improve the suppression and control of $\alrb$'s.  An overview of the polarized source laser and optics systems as they are configured for $\text{E-158}$ is illustrated in Figure~\ref{fig:overview}. The laser and optics systems are housed in an environmentally controlled room outside of the accelerator tunnel.  The ``Flash:Ti Bench'' holds the laser cavity and pulse-shaping optics. The ``Diagnostics Bench'' has photodiodes for monitoring the laser's intensity and temporal profile and a monochromator for measuring its wavelength. The ``Helicity Control Bench'' houses the optics for controlling the polarization state of the beam and for suppressing $\alrb$'s. A 20-m Transport Pipe transports the beam into the accelerator tunnel, where it crosses the ``Cathode Diagnostics Bench'' and is directed onto the cathode of the polarized gun. The ``Cathode Diagnostics Bench'' holds optics for setting the position of the beam spot on the cathode and an auxiliary diagnostic line.  The photoelectrons emitted by the cathode are bent through 38$^{\circ}$ and enter the accelerator.  
\begin{longtable}{|l|c|}
\caption{Parameters of the Flash:Ti laser beam as it ran for $\text{E-158}$ 2002 Physics Run I.  The position jitter at the photocathode is measured on the electron beam but is dominated by laser jitter.  The other entries are measured directly on the laser beam.\label{tab:flashti}} \\
\hline
\hline
\rule[-1ex]{0pt}{3.5ex}  Wavelength & 805 nm\footnote{Tunable over $750-850\ \text{nm.}$} \\
\hline
\rule[-1ex]{0pt}{3.5ex}  Bandwidth & $0.7\ \text{nm FWHM}$ \\
\hline
\rule[-1ex]{0pt}{3.5ex}  Repetition rate & 120 Hz  \\
\hline
\rule[-1ex]{0pt}{3.5ex}  Pulse length & 270 ns\footnote{Tunable over $50-370\ \text{ns.}$} \\
\hline
\rule[-1ex]{0pt}{3.5ex}  Pulse energy & 60 $\mu$J\footnote{Typical operating energy.  The maximum available energy is $600\ \mu\text{J}$ in a $370\text{-ns}$ pulse.}  \\
\hline
\rule[-1ex]{0pt}{3.5ex}  Circular polarization & 99.8\ \%  \\
\hline
\rule[-1ex]{0pt}{3.5ex}  Energy jitter & $0.5\ \%\ \text{rms}$  \\
\hline
\rule[-1ex]{0pt}{3.5ex}  Position jitter at photocathode & $<\ 70\ \mu\text{m rms}$\footnote{For $1\ \text{cm}$ 1/e$^2$ diameter.} \\
\hline
\hline
\end{longtable}

\begin{figure}
   \centering
   \begin{tabular}{c}
   \includegraphics{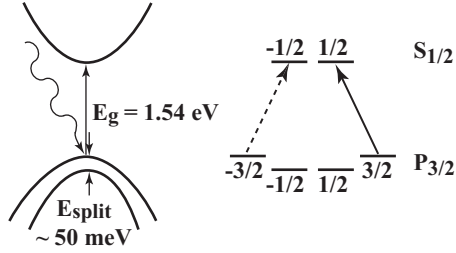}
   \end{tabular}
   \caption[example] 
   { \label{fig:GaAslevels} 
A diagram of the bandgap and energy levels for strained GaAsP.  The arrows indicate the allowed transitions for right- and left-helicity photons of $\lambda = 805\ \text{nm}$.  Adapted from~\cite{saez}.}
   \end{figure}

The heart of the polarized electron source is its photocathode.  A new gradient-doped strained GaAsP cathode was installed prior to \e158's 2002 physics running.  This cathode, which is more fully described in~\cite{takashi}, was developed in a R\&D project for the Next Linear Collider (NLC) project~\cite{nlcbook}.  NLC requires roughly 2.5 times more charge than \e158 in a $270\text{-ns}$ pulse and $>\ \sim\!80\ \%$ electron polarization~\cite{turner}.  With the available laser power, this cathode can yield a charge of $2\cdot 10^{12}$ electrons in $100\ \text{ns}$.  This is significantly more charge than is required by \e158 (and significantly more than yielded by the previous cathode), providing additional flexibility in optimizing the optics system.  In order to provide an electron beam polarization as high as $\sim 80\ \%$, a strain is applied to the active layer of the cathode to break the degeneracy of the P$_{3/2}$ energy levels as illustrated in Figure~\ref{fig:GaAslevels}.  However, the amount of strain varies in direction with respect to the crystalline lattice.  This variation induces a ``QE anisotropy'' in the cathode, whereby the quantum efficiency (QE) of the cathode becomes dependent on the orientation of the linear polarization of incident laser light, with a typical analyzing power of $5-15\ \%$~\cite{Mair96}.  The QE anisotropy is a dominant ingredient contributing to $\alrb$'s and its effects are discussed in section 4.  

\begin{figure}
   \centering
   \begin{tabular}{c}
   \includegraphics{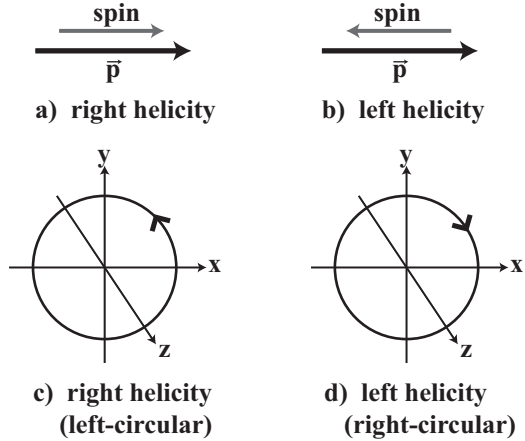}
   \end{tabular}
   \caption[example] 
   { \label{fig:defn} 
Conventions used for electron beam helicity and laser beam polarization.  a) and b) define right- (left-) helicity photon and electron beams as having the particle spin aligned (antialigned) with the particle momentum.  c) and d) show the commonly used optics convention.  The arrows on the polarization ellipses indicate increasing time.}
   \end{figure}

The relationship between the helicity of the laser beam and the helicity of the resulting polarized electron beam can be determined by considering the bandgap diagram for GaAsP, shown in Figure~\ref{fig:GaAslevels}.  The laser light pumps electrons from the $P_{3/2}$ valence band into the $S_{1/2}$ conduction band.  Right-helicity laser light excites electrons into the $m_J=-1/2$ state in the conduction band.  Because we operate the cathode in reflection mode (the emitted electrons move in the direction opposite of the incoming laser light), the extracted electrons are also right-helicity.  Similarly, left-helicity laser light excites electrons into the $m_J=+1/2$ state, and in reflection yields left-helicity electrons.  In this paper we define the handedness of the photon and electron beams according to their helicity~\cite{ql} as shown in Figure~\ref{fig:defn}.\footnote{This convention results in right- (left-) helicity photons corresponding to left- (right-) circular polarization photons in the commonly used optics convention given in~\cite{bw}.}


\begin{figure}
   \centering
   \begin{tabular}{c}
   \includegraphics{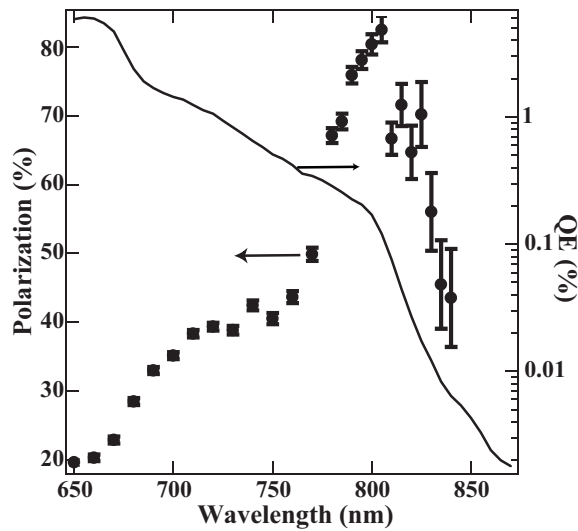}
   \end{tabular}
   \caption[example] 
   { \label{fig:polqe} 
Electron polarization and cathode QE as a function of wavelength.  This data is from a different sample of the same GaAsP wafer as the cathode installed prior to \e158 physics running.  Reprinted from~\cite{takashi}.}
   \end{figure}

The operating wavelength and required bandwidth of the Flash:Ti are determined by the photoemission properties of the cathode.  Figure~\ref{fig:polqe} shows, as a function of wavelength, the cathode QE and the polarization of the photoemitted electrons for the cathode used during $\text{E-158}$ 2002 Physics Run I.  As shown in Table~\ref{tab:flashti}, we have chosen an operating wavelength of $805\ \text{nm}$ in order to sit at the polarization peak, and the Flash:Ti's $0.7\ \text{nm FWHM}$ bandwidth is narrow enough to ensure that all of the laser power is used to generate electrons at the peak polarization.

A particular focus of this paper is on the parts of the optical system that are responsible for producing a circularly polarized laser beam and for suppressing and controlling $\alrb$'s.  As is discussed in section~\ref{sec:pita}, a high degree of circular polarization is critical for minimizing $\alrb$'s.  We use a linear polarizer and a pair of Pockels cells on the Helicity Control Bench to gain control over the laser beam polarization and to switch between helicity states on a pulse-by-pulse basis.  To achieve further suppression of $\alrb$'s, a number of features are built into the optical system.  These features include \\
\begin{enumerate}{}{\setlength{\topsep}{6pt}\setlength{\itemsep}{6pt}\setlength{\parsep}{0pt}\setlength{\parskip}{0pt}}
\item [i)] optimization of the beam spot size and waist location at the polarization Pockels cells; \\
\item [ii)] imaging of the polarization Pockels cells onto the photocathode; \\
\item [iii)] active feedbacks to null $A_I$, $D_{X(Y)}$, and $D_{X'(Y')}$; \\
\item [iv)] an insertable half-wave plate to reverse the laser helicity; and \\
\item [v)] the ability to toggle between two beam expanders which provide a magnification of equal magnitude and opposite sign to reverse certain contributions to $D_{X(Y)}$ and $D_{X'(Y')}$.
\end{enumerate}

This paper includes performance results for the laser and optics system from three runs:  $\text{T-437}$, a test-beam run in November 2000 which commissioned the polarized source for \e158; an \e158 engineering run in January-May 2001; and \e158 Physics Run I in April-May 2002.  Table~\ref{tab:runs} summarizes changes in key parameters of the laser beam between runs.  The wavelength was changed prior to 2002 Physics Run I to accomodate the new cathode.  In addition, a few results are presented from the Gun Test Laboratory, a test facility which reproduces the first few meters of the beam line.  
\begin{table}[h]
\caption{Summary of changes in key operating parameters of the Flash:Ti laser between $\text{E-158}$-related runs.} 
\label{tab:runs}
\begin{center}       
\begin{tabular}{|l|c|c|c|} 
\hline
\hline
\rule[-1ex]{0pt}{3.5ex} Run & Laser Rate & Wavelength & Energy Jitter   \\
\hline
\rule[-1ex]{0pt}{3.5ex} $\text{T-437}$ & 60 Hz & 852 nm & 1.0 \%   \\
\hline
\rule[-1ex]{0pt}{3.5ex} 2001 Engineering & 60 Hz & 852 nm & 1.5 \%   \\
\hline
\rule[-1ex]{0pt}{3.5ex} 2002 Physics Run I & 120 Hz & 805 nm & 0.5 \%   \\
\hline
\hline
\end{tabular}
\end{center}
\end{table}

Sections 2 and 3 of this paper discuss the design and implementation of the Flash:Ti laser system and the accompanying optics system, progressing in sequence from the laser cavity to the photocathode.  Section~\ref{sec:pita} discusses how $\alrb$'s can arise from interactions between imperfections in the laser circular polarization and the cathode's QE anisotropy.  Section 5 discusses how the laser polarization and transport optical systems are configured and optimized to suppress $\alrb$'s and presents results from $\text{T-437}$.  Finally, section 6 summarizes many effects which can generate $\alrb$'s and notes their relevance to the SLAC source and $\text{E-158}$.

%% file: n2.tex
\section{$\mathbf{The}$ $\mathbf{Flash\!:\!Ti}$ $\mathbf{Laser}$ $\mathbf{System}$}
\setcounter{footnote}{0}

This section discusses the design and operation of the Flash:Ti laser cavity,
the Flash:Ti cooling flow system, the Flash:Ti modulator, the pulse shaping and intensity control optics, and the laser beam diagnostics. The Flash:Ti laser was largely designed and built at SLAC and is unique for its low jitter, long pulse length, and high repetition rate capability.  Performance results from the recent $\text{E-158}$ engineering and physics runs are presented.

\subsection{Flash:Ti Laser Cavity}
\begin{figure}
   \centering
   \begin{tabular}{c}
   \includegraphics{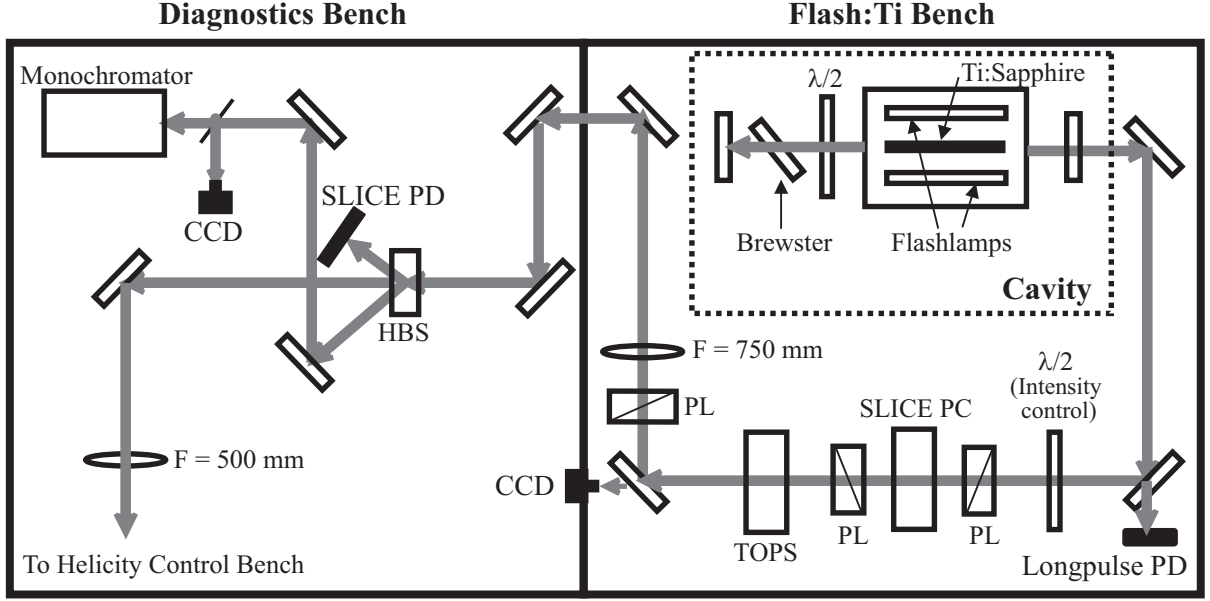}
   \end{tabular}
   \caption[example] 
   { \label{fig:ft} 
Schematic of the laser cavity and the optical layout of the Flash:Ti and Diagnostics Benches ($\lambda/2$: half-wave plate, PD: photodiode, PL: polarizer, PC: Pockels cell, HBS: holographic beam sampler).}
   \end{figure}

The Flash:Ti pump chamber was designed at SLAC and constructed for us by Big Sky Laser Technologies.\footnote{Big Sky Laser Technologies, Bozeman, MT, USA.}  A schematic of the laser cavity is depicted in Figure~\ref{fig:ft}.  The rod-shaped Ti:Sapphire crystal is pumped by two flashlamps, each of which is associated with an elliptically shaped reflector. The original commercial silver coatings of the reflectors and pump chamber end plates have been replaced by rhodium.  This change substantially increased their mechanical and chemical surface durability and eliminated the need to purge the pump chamber with nitrogen during flashlamp changes or other maintenance work.  The pump chamber parts can be exposed to air while they are handled without the risk of corrosion. Two cylindrical flashlamps\footnote{Model  L8061E, T J Sales Associates Inc., ILC Technology Inc., Denville, NJ, USA.} are used for this system.  The flashlamp tubes have the following specifications: ID $4.8\ \text{mm}$, OD $5.98\ \text{mm}$, $7.6\text{-inch}$ arc length, Ce-doped quartz walls, and $450\ \text{Torr}$ Xe filling. The output of the flashlamps is focused on the center of a $4\text{-mm}$ diameter $0.1 \%\text{-doped}$ Ti:Sapphire laser rod\footnote{Union Carbide Crystal Products, Washougal, WA, USA.} of $6.4\text{-inch}$ length.  The rod, flashlamps, and pump chamber are cooled by a closed loop of ultra-pure water. The rod flow tube\footnote{$KTF\text{-2}$ fluorescent converter, Kigre, Inc., Hilton Head, SC, USA.} surrounds the laser rod and its material acts as a UV filter to prevent excessive solarization of the Ti:Sapphire material.  

We achieve maximum laser cavity output power while maintaining low pulse-to-pulse jitter by using a one-meter-long cavity formed by an $85\ \%\text{-reflectivity}$ planar output coupler and a $99.9\ \%\text{-reflectivity}$ end mirror with a $2\text{-m}$ concave curvature.  Both mirrors have narrow-band dielectric coatings centered at the operating wavelength ($800\ \text{nm}$ or $850\ \text{nm}$). A single quartz quarter-wave plate of $\sim1.3\ \text{mm}$ thickness acts as both a Brewster plate and a birefringent tuner. It is mounted to allow for both horizontal rotation and rotation about the axis normal to its surface. In the horizontal plane the plate is set to the Brewster angle of $\sim57$\textdegree. The effective refractive index of the quartz plate depends on the angle between the electric field vector and the optical orientation of the quartz plate. We achieve birefringent wavelength tuning by rotating the quartz plate about the axis normal to its surface.  This optimizes the transmission for the desired output wavelength of $805\ \text{nm}$ ($852\ \text{nm}$ for $\text{T-437}$ and the 2001 engineering run) with a bandwidth of $\sim0.7\ \text{nm}$ (FWHM).  A half-wave plate located between the laser head and the Brewster plate compensates for the arbitrary orientation of the Ti:Sapphire laser rod and thereby guarantees that p-polarization transmission is maximized through the Brewster plate. Recent modifications of the laser head assembly procedure allow installation of the laser rod with control of its crystallographic orientation. This eliminates the need for the half-wave plate inside the cavity and further improves the performance of the cavity stability. Preliminary measurements in our development laboratory indicate a pulse-to-pulse jitter of $\sim0.3\ \%\ \text{rms}$. The equivalent modification of the laser head used at the polarized electron source is planned for the next $\text{E-158}$ physics run. 

\subsubsection{Thermal Lensing}
\label{sec:tlens}
\begin{figure}
   \centering
   \begin{tabular}{c}
   \includegraphics{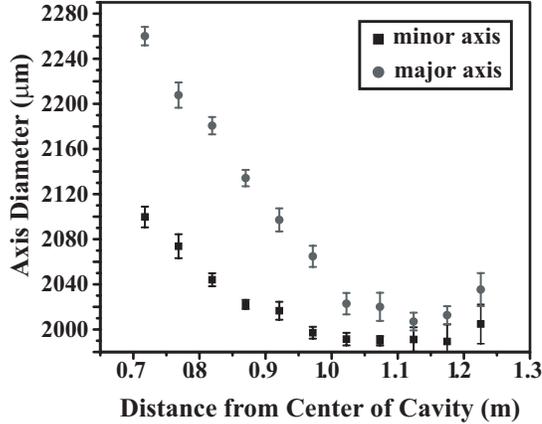}
   \end{tabular}
   \caption[example] 
 { \label{fig:thermlens} 
Focusing of the laser beam due to thermal lensing at a laser wavelength of 805 nm and a flashlamp voltage of $8\ \text{kV}$.  The dimensions of the laser spot ellipse are measured at the $1/e^2$ level.
}
   \end{figure}

Pumping the Ti:Sapphire rod with flashlamps leads to a strong thermal lensing effect~\cite{siegman}.  To investigate the power of the thermal lens for our system, the beam spot has been analyzed at relevant locations along the beam path under typical running conditions. A set of measurements is shown in Figure~\ref{fig:thermlens}. The lengths of the minor and major axes of the ellipse formed by the laser spot decrease with distance from the cavity center and increase again after the focal waist has been reached. The measurements indicate a focus at $\sim1.1\ \text{m}$ from the center of the cavity. Compared to the curvature of the resonator mirrors, the thermal lens is the dominating optic. To optimize the laser stability, thermal lensing has been considered for end mirror selection and mirror spacing.

\subsubsection{Flash:Ti Cooling Flow System}
\begin{figure}
   \centering
   \begin{tabular}{c}
   \includegraphics{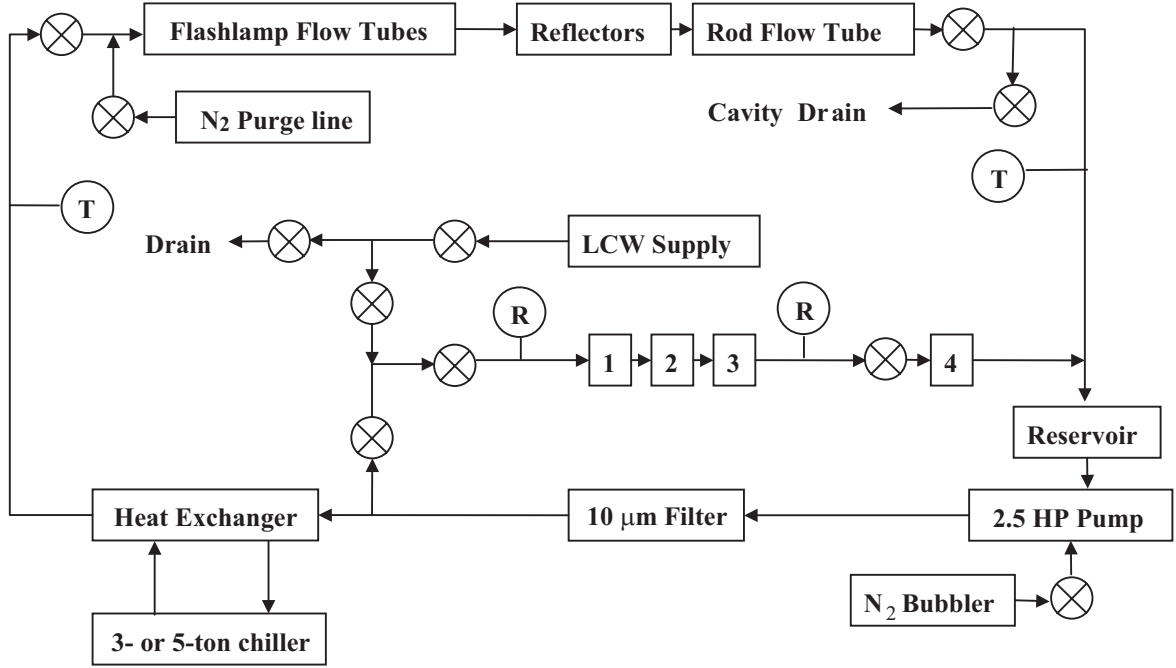}
   \end{tabular}
   \caption[example] 
 { \label{fig:flow} 
Cooling water system components and water flow, including resistivity sensors (\normalsize{$\bigcirc$}\small \letterspace to -1.6\naturalwidth{R}\hspace{0.16in}), temperature sensors (\normalsize{$\bigcirc$}\small \letterspace to -1.6\naturalwidth{T}\hspace*{0.16in}), valves ($\bigotimes$), on-site low conductivity water (LCW), oxygen removal cartridge (1), activated carbon-organics filter (2), mixed bed deionizer (3), and submicron filter (4).
}
   \end{figure}

The cooling water flow system is a closed loop and can be refilled from on-site low-conductivity water. Figure~\ref{fig:flow} shows a schematic of the water flow system. The system is designed to ensure a constant temperature of $76\text{\textdegree\ F}$ and nearly inert water conditions.  The main loop is chilled by a heat exchanger and operates at a flow rate of $2.5\ \text{GPM}$. Ultra-pure water quality ( $> 15\ \text{M}\Omega$) is established by a $1\text{-GPM}$ polishing loop which contains a $0.2\ \mu\text{m}$ particle filter, a deionization filter, an organic filter, and an oxygen filter. A nitrogen bubbler significantly reduces the partial pressure of oxygen in the reservoir, minimizing the amount of oxygen dissolved in the water. This was of particular importance when the silver-coated reflectors were in use. The cooling water constantly flows through a $10\text{-}\mu\text{m}$ particle filter and a heat exchanger. Either a $3\text{-ton}$ or a $5\text{-ton}$ chiller can be used to provide cooling for the heat exchanger. Interlocks are connected to water flow, resistivity, and temperature sensors mounted near the laser head.  The interlocks shut off the laser power supply if the sensor values move out of tolerance. For flashlamp changes the laser head can be drained and purged by a separate N$_2$ supply.

\subsubsection{Flash:Ti Modulator}

\begin{figure}
   \centering
   \begin{tabular}{c}
   \includegraphics{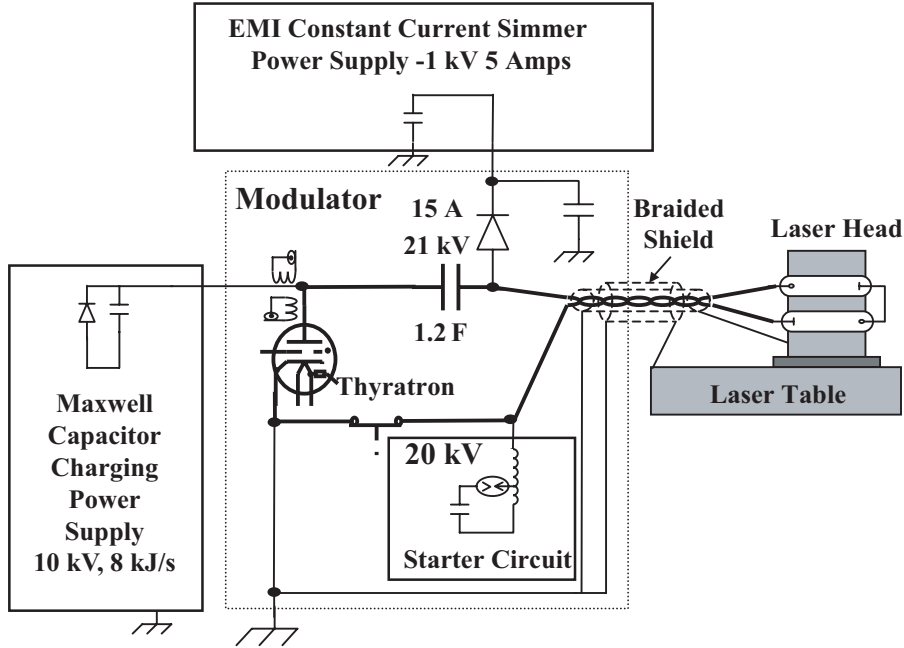}
   \end{tabular}
   \caption[example] 
   { \label{fig:mod} 
Schematic circuit diagram of the Flash:Ti power supply.}
   \end{figure}

The modulator (Figure~\ref{fig:mod}) was designed and built by SLAC personnel and provides the high voltage pulse needed to fire the flashlamps.  A 1.2 $\mu$F capacitor charges from a $10\ \text{kV}$, $8\ \text{kJ/s}$ power supply.\footnote{CCDS 810TI, Maxwell Technologies, San Diego, CA, USA.} Upon ignition of a thyratron, the capacitor discharges through the two flashlamps in series. This produces an over-damped electrical pulse whose characteristics are set by the capacitance of the capacitor and the stray inductance and resistance of the circuit components. The pulse has a peak current of $1\ \text{kA}$ and a duration of $22\ \mu\text{s}$. Between pulses, a current through the flashlamps is maintained by a ``simmer'' power supply.\footnote{Model 1000TS, EMI, Neptune, NJ, USA.} The simmer current reduces the high voltage needed for conduction in the lamps and thereby extends their lifetime. 

\subsection{Intensity Control and Pulse Shaping}
\label{sec:TOPS}
Immediately following the laser cavity on the Flash:Ti bench in Figure~\ref{fig:ft} are optics dedicated to controlling the laser pulse's energy, length, and temporal profile.  Located between a pair of crossed polarizers, the ``SLICE'' Pockels cell is used to control the laser pulse's energy and pulse length.  The ``start'' trigger for the SLICE Pockels cell is set at the beginning of the low-jitter section of the laser pulse (see Figure~\ref{fig:beam}b). The duration of the sliced pulse is set by its ``stop'' trigger. Typical sliced pulse lengths are $50-370\ \text{ns}$. The SLICE Pockels cell is driven by a commercial high voltage pulser.\footnote{Model $PVX\text{-3110}$, Directed Energy, Inc., Fort Collins, CO, USA.} The amplitude of the high voltage pulse controls the intensity of the laser pulse. We use the SLICE amplitude as the control device of a feedback system to stabilize the intensity of the electron beam.  This feedback provides compensation for the slow decrease in cathode QE during its $3\text{-day}$ cesiation cycle as well as the slow degradation of the flashlamps' efficiency during their lifetime.  The half-wave plate located upstream of the SLICE Pockels cell provides a means of limiting the maximum sliced laser power to a level that is safe for accelerator operation.

\begin{figure}
   \centering
   \begin{tabular}{c}
   \includegraphics{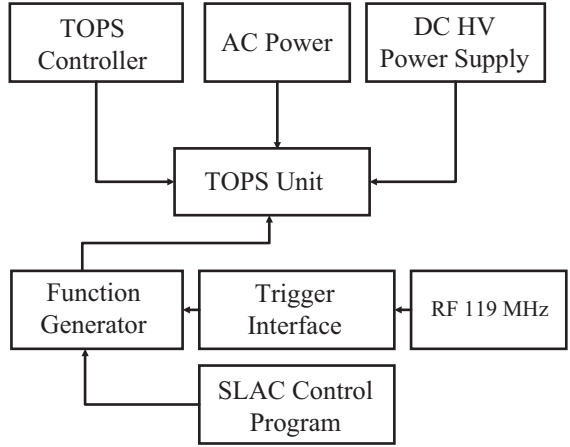}
   \end{tabular}
   \caption[example] 
   { \label{fig:TOPS} 
Schematic of the TOp-hat Pulse Shaper electronics.  The TOPS unit is installed on the Flash:Ti Bench (see Figure~\ref{fig:ft}) and contains the TOPS Pockels cell.}
   \end{figure}

We shape the laser pulse's temporal profile using a Pockels cell--polarizer pair (TOPS, TOp-hat Pulse Shaper, shown schematically in Figure~\ref{fig:TOPS}) installed downstream of the SLICE Pockels cell. This shaping is used to compensate for beam loading and to achieve a small energy spread on the electron beam as described below in section~\ref{sec:bl}.  TOPS is a SLAC-built pulse-shaping Pockels cell system driven by a Stanford Research Systems (SRS) DS345 $30\ \text{MHz}$ synthesized function generator. The SRS DS345 synthesized function generator has been modified internally.  It uses the SLAC $119\text{-MHz}$ source as an oscillator.  Power for the Pockels cell is supplied by a SLAC-built DC power supply.  Control power for the TOPS system is supplied by the TOPS controller unit.  The reference signal, which controls the Pockels cell, comes from the function generator. The function generator is integrated into the SLAC Control Program (SCP) through its GPIB interface.  Remote control is achieved via an EPICS (Experimental Physics and Industrial Control System) user interface.  The function generator allows one to generate an arbitrary waveform in $25\text{-ns}$ steps.  Using TOPS to compensate for beam loading and minimize each pulse's energy spread is discussed further in section~\ref{sec:bl}.

\subsection{Diagnostics}
The laser beam is folded at multiple locations using broadband NIR-coated (near-infrared) high-reflectivity mirrors. We use leakage light through these mirrors or a sampled beam for diagnostic purposes. We routinely monitor laser intensity, jitter, wavelength and spot size. The locations of the diagnostics are shown in Figure~\ref{fig:ft}. One photodiode installed upstream of the pulse-shaping optics monitors the Flash:Ti laser output (Longpulse PD).  A holographic beam sampler\footnote{Gentec Electro-Optics, Quebec, QC, Canada.} downstream of the pulse-shaping optics supplies two one-percent samples of the laser beam. One sample is used to monitor the intensity of the sliced pulse (SLICE PD). The second sample is focused onto a scanning monochromator for wavelength measurements or can be used to image the beam spot onto a CCD camera.

\subsection{Laser Performance}

We summarize below the performance of the upgraded Flash:Ti laser system during recent running.  We briefly review the laser's performance during \t437 and the 2001 engineering run, and then focus on its performance during 2002 Physics Run I.  The performance of the laser system for the earlier runs is more fully described in~\cite{brachmann}.  The operating parameters of the laser system for 2002 Physics Run I are summarized above in Table~\ref{tab:flashti}.

\subsubsection{Flash:Ti Performance During \t437 and the \e158 Engineering Run}
\t437 and the \e158 2001 engineering run preceded the Flash:Ti cavity optimization and thermal lensing studies.  In addition, the cathode used for those runs required a wavelength of $852\ \text{nm}$ for maximum electron polarization, causing the laser to operate fairly far from the gain maximum for Ti:Sapphire.  We achieved a laser power of $\sim 20\ \text{mJ}$ in a $15\text{-}\mu\text{s}$ laser pulse with the laser cavity tuned to this wavelength.  The SLICE Pockels cell described in section~\ref{sec:TOPS} was set to slice a $130\text{-ns}$ pulse ($370\ \text{ns}$ for \t437) out of the area of highest stability, resulting in a pulse of $1.5\ \%$ rms intensity jitter ($1.0\ \%$ rms for \t437).  The pulse energy was $\sim 175\ \mu\text{J}$ for these conditions during the engineering run.

\subsubsection{Flash:Ti Performance During \e158 2002 Physics Run I}
A number of modifications improved the performance of the laser system for \e158 2002 Physics Run I.  First, the new photocathode requires a laser wavelength of $805\ \text{nm}$ for peak electron polarization. At $805\ \text{nm}$ the laser operates closer to the gain maximum of the Ti:Sapphire laser crystal, yielding a significant enhancement of laser performance. Furthermore, the consideration of thermal lensing described in section~\ref{sec:tlens} and appropriate end mirror selection were essential for improved performance.  We also began to study the dependence of energy jitter on the laser power supply high voltage and the current of the switching thyratron. These were then optimized to minimize the laser's energy jitter.  As the flashlamps and thyratron age they require small adjustments of the thyratron reservoir voltage.  

\begin{figure}
   \centering
   \begin{tabular}{c}
   \includegraphics{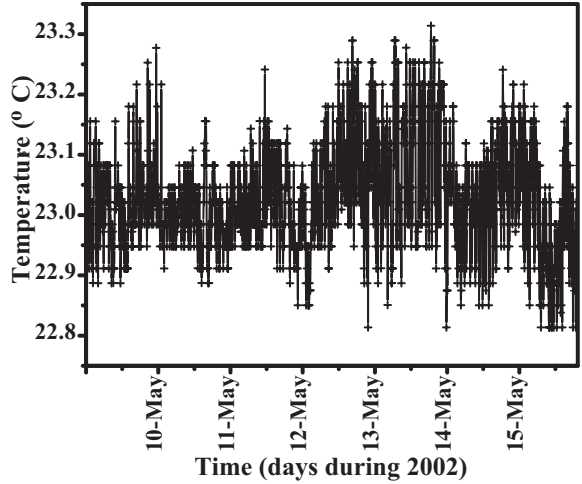}
   \end{tabular}
   \caption[example] 
   { \label{fig:temp} 
Temperature stability of the Laser Room housing the polarized source laser and optics systems over a one-week time period.}
   \end{figure}

\begin{figure}
   \centering
   \begin{tabular}{c}
   \includegraphics{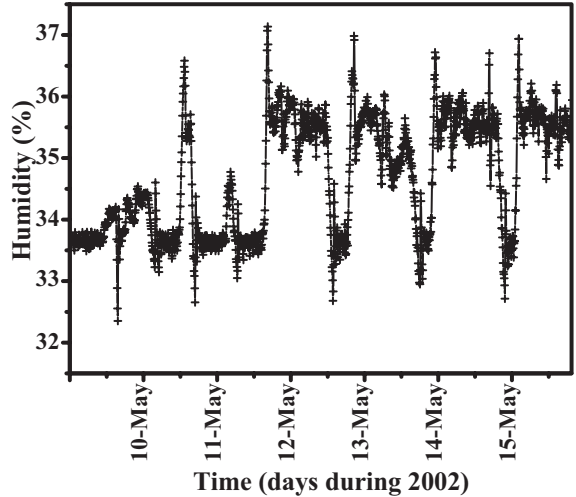}
   \end{tabular}
   \caption[example] 
   { \label{fig:humi} 
Humidity stability of the Laser Room housing the polarized source laser and optics systems over a one-week time period.}
   \end{figure}

Slow drifts in laser power and stability caused by humidity and temperature variations are minimized by appropriate air conditioning and humidity control. The temperature and humidity in the laser room are stable at $23.0 \pm 0.1\ \text{\textdegree C}$ and $35 \pm 1\ \%$, respectively.  Figures~\ref{fig:temp} and~\ref{fig:humi} show the stability of the temperature and humidity in the Laser Room which houses the polarized source laser and optics systems.

\begin{figure}
   \centering
   \begin{tabular}{c}
   \includegraphics{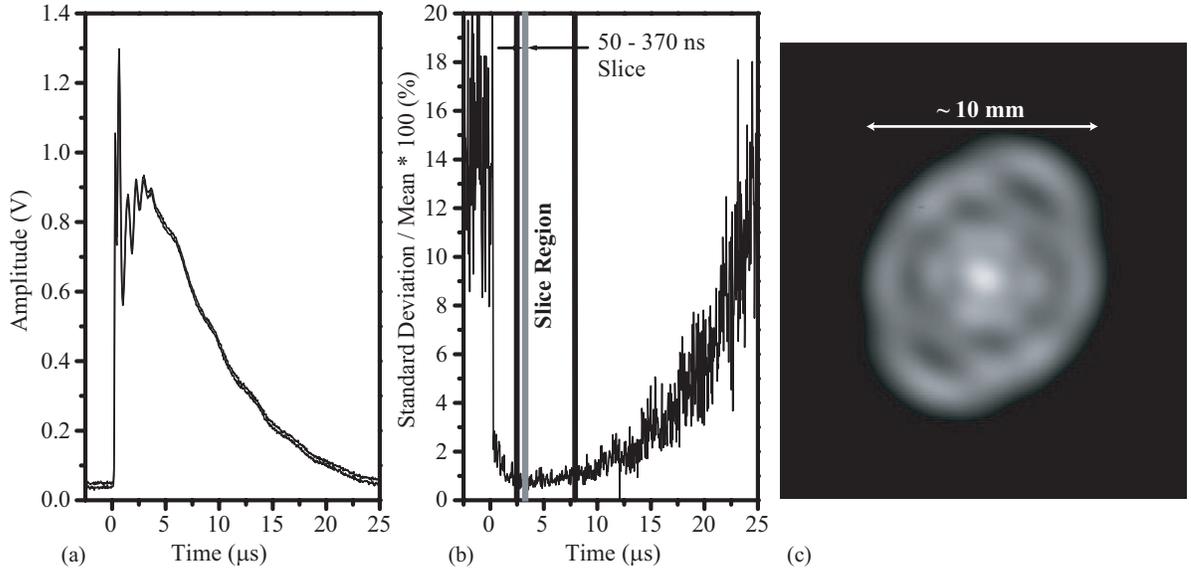}
   \end{tabular}
   \caption[example] 
   { \label{fig:beam} 
(a) Photodiode signal of the Flash:Ti laser pulse.  The two traces form an envelope about a sample of 100 measured pulses, indicating the pulse-to-pulse stability of the laser's temporal profile.  (b) Energy jitter as a function of time within the pulse.  The SLICE Pockels cell is used to select the $50-370\ \text{ns}$ region of lowest energy jitter, marked by the vertical lines.  (c) The spatial profile of the sliced pulse.}
   \end{figure}

\begin{figure}
   \centering
   \begin{tabular}{c}
   \includegraphics{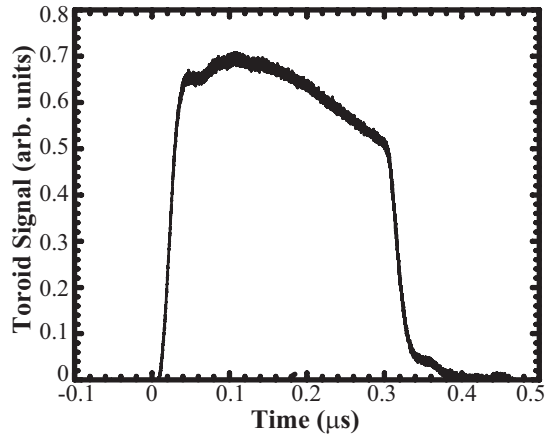}
   \end{tabular}
   \caption[example] 
   { \label{fig:topsshape} 
Electron beam temporal profile resulting from using TOPS for beam-loading compensation.}
   \end{figure}

Typical cavity output power at $805\ \text{nm}$ is $\sim45\ \text{mJ}$. Pulse slicing provides a pulse of $50-370\ \text{ns}$ with a maximum energy of $\sim600\ \mu\text{J}$ per pulse (in $370\ \text{ns}$).  During typical physics running, the laser pulse provided $\sim 60\ \mu\text{J}$ in $270\ \text{ns}$ in order to generate an electron beam pulse of $\sim 6 \cdot 10^{11}\ \text{electrons/spill}$.  Figures~\ref{fig:beam}a and~\ref{fig:beam}b show the temporal shape and the stability of the $15\text{-}\mu\text{s}$ laser pulse for a 100-pulse sample.  Also indicated in Figure~\ref{fig:beam}b is the area of slicing, located at the point in time at which the laser energy jitter is at a minimum. The spatial profile of the laser beam, measured on the Diagnostics Bench and shown in Figure~\ref{fig:beam}c, indicates the multimodal structure of the laser pulse.  Multimodal operation of the laser is necessary in order to generate the $15\text{-}\mu\text{s}$ pulse from which $50-370\ \text{ns}$ can be sliced with a flat-top profile.  Figure~\ref{fig:topsshape} shows the temporal profile of the electron beam at the first fast toroid following the cathode.  Its profile reflects the profile of the sliced laser beam after it has been shaped by TOPS in order to compensate for beam-loading effects.

\begin{figure}
   \centering
   \begin{tabular}{c}
   \includegraphics{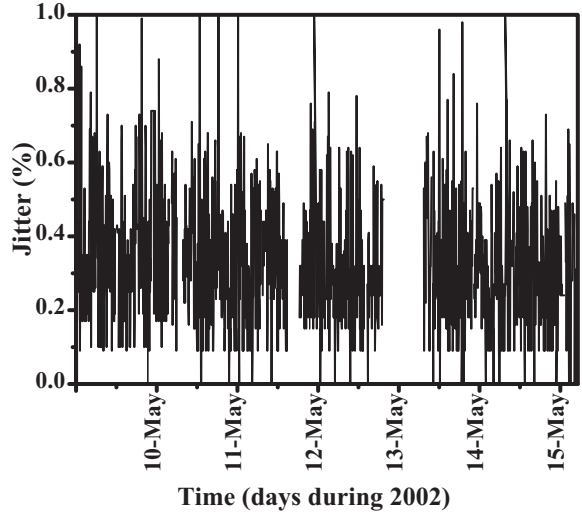}
   \end{tabular}
   \caption[example] 
   { \label{fig:t488j} 
Intensity jitter of the electron beam at the first toroid following the cathode over a one-week time period.  Each point on the plot represents the rms of 10 sampled pulses.}
   \end{figure}

\begin{figure}
   \centering
   \begin{tabular}{c}
   \includegraphics{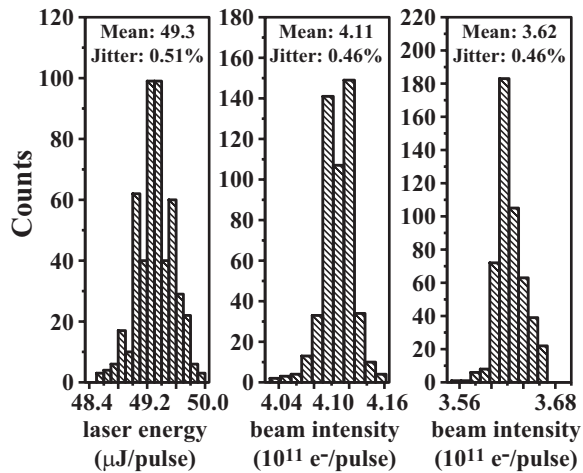}
   \end{tabular}
   \caption[example] 
   { \label{fig:histo} 
(a) Histogram of the energy jitter in the sliced laser pulse.  (b) Histogram of the intensity jitter in the electron beam in the injector.  (c)  Histogram of the intensity jitter in the electron beam at the end of the accelerator.}
   \end{figure}

\begin{figure}
   \centering
   \begin{tabular}{c}
   \includegraphics{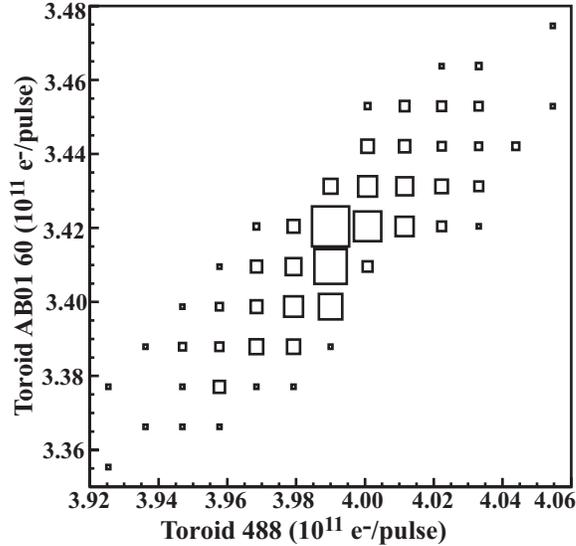}
   \end{tabular}
   \caption[example] 
   { \label{fig:torcorr} 
Correlation plot of raw intensities as measured by toroid 488, located in the injector, and toroid AB01 60, located between the Linac and the \e158 target.  The size of each square is proportional to the number of counts in its bin.}
   \end{figure}

A pulse stability of $\sim0.5\ \%$ rms was maintained throughout the run.  A time history of the intensity jitter in the electron beam for a typical one-week period as measured by the first toroid downstream of the cathode (toroid 488) is shown in Figure~\ref{fig:t488j}.  The stability of the laser and electron beam intensities at $120\ \text{Hz}$ as measured at the SLICE photodiode and toroid 488 are shown in Figures~\ref{fig:histo}a and~\ref{fig:histo}b, respectively.  Figure~\ref{fig:histo}c shows the pulse-to-pulse jitter of the polarized electron beam at toroid AB01 60 located at the end of the accelerator near the target. The high degree of correlation between toroid 488 and toroid AB01 60, shown in Figure~\ref{fig:torcorr}, demonstrates the importance of a highly stable electron source. Almost no additional instabilities in the intensity are introduced throughout the two-mile-long accelerator.

The maintenance of the laser system during the \e158 runs ($5\ \text{months}$ continuous operation at $60\ \text{Hz}$ for the engineering run and $5\ \text{months}$ for the physics run at a mix of $60\ \text{Hz}$ and $120\ \text{Hz}$) consists of flashlamp changes every $\sim1.45 \cdot 10^8$ laser pulses (28 days at $60\ \text{Hz}$ and 14 days at $120\ \text{Hz}$) and cooling water system filter changes every 6 months.  Changing the flashlamps requires one hour and can often be scheduled to occur during other planned interruptions to beam delivery, making the impact of laser system maintenance on \e158's running efficiency negligible.  We observe no significant drop in laser performance due to the aging of flashlamps or water filters.

%% file: n3.tex
\section{$\mathbf{Helicity}$ $\mathbf{Control}$ $\mathbf{Bench}$ $\mathbf{and}$ $\mathbf{Transport}$ $\mathbf{Optics}$ $\mathbf{to}$ $\mathbf{Cathode}$}
\setcounter{footnote}{0}
\label{sec:polcon}
In this section we describe the optics that follow the laser and pulse-shaping systems.  These optics circularly polarize the laser beam in a manner that permits selecting the helicity in a pseudorandom sequence on a pulse-by-pulse basis.  The circularly polarized light is then transported to the cathode and care is taken to minimize distortions in the laser polarization.  As mentioned below (and discussed in detail in section~\ref{sec:pita}), small distortions generate significant amounts of linear polarization and ultimately contribute to helicity-correlated asymmetries.  The optics are configured to passively minimize and actively null $\alrb$'s.

\subsection{Circular Polarization}
\label{sec:wb}
\begin{figure}
   \centering
   \begin{tabular}{c}
   \includegraphics{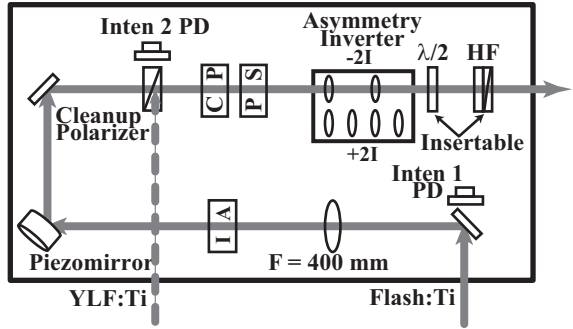}
   \end{tabular}
   \caption[example] 
   { \label{fig:hc} 
The Helicity Control Bench contains the optics for control of the laser beam's polarization and $\alrb$'s.}
   \end{figure}

The polarization optics, shown in Figure~\ref{fig:hc}, are designed to generate highly circularly polarized light of either helicity while minimizing $\alrb$'s.
The ``Cleanup Polarizer" and the ``Circular Polarization'' (CP) and ``Phase Shift'' (PS) Pockels cells collectively determine the polarization of the beam.\footnote{All three Pockels cells on the Helicity Control Bench (CP, PS, and IA) are $20\text{-mm}$-aperture Cleveland Crystals model QX2035, specified to be windowless, to be parallel to better than 0.5 arcminute, and to be broadband AR-coated centered at $800\ \text{nm}$.  Cleveland Crystals, Inc., 676 Alpha Dr., Highland Hts., OH 44143, USA.}  The Cleanup Polarizer also functions to combine the YLF:Ti beam used to generate electrons for the PEP rings\footnote{For \e158's 2002 Physics Run I, it shared accelerator pulses with the BaBar experiment that utilized the PEP storage rings~\cite{decker}.  The BaBar experiment does not utilize the beam polarization.} with the Flash:Ti beam so that they share a common path through the remaining transport optics.  The CP cell acts as a quarter-wave plate with its fast axis at 45\textdegree\ from the horizontal. The sign of its retardation can be chosen on a pulse-by-pulse basis, generating circularly polarized light of either helicity. Its quarter-wave voltage is approximately $2.7\ \text{kV}$. Adjusting its voltage from the quarter-wave setting allows the CP cell to compensate for linear polarization along the horizontal and vertical axes that arises from either residual birefringence in the Pockels cell or phase shifts in the optics between the Pockels cells and the photocathode. The PS cell, with a vertical fast axis, is pulsed at low voltages ($\lesssim150\ \text{V}$) and is used to compensate for residual linear polarization along the axes at $\pm45$\textdegree.  The phase shift $\delta_{CP (PS)}$ induced by the CP (PS) cell is given by the relation
\begin{equation}
\delta_{CP (PS)} = \frac{\pi}{V_{\lambda/2}} \cdot V_{CP (PS)},
\label{eq:mradvolts}
\end{equation}
where $V_{\lambda/2}$ is the voltage required for half-wave retardation (typically $\sim5.4\ \text{kV}$) and $V_{CP (PS)}$ is the voltage across the CP (PS) cell.

In the following subsections we discuss the polarization control given by the CP and PS cells, our procedures for aligning them, the insertable half-wave plates that are used to generate a slow helicity reversal, and the data acquisition and control systems used to determine the polarization sequence and set the Pockels cell voltages.

\subsubsection{Polarization Analysis}
To understand the function of the CP and PS cells, it is useful to develop expressions for the polarization of the laser beam in terms of the phase shifts induced by the CP and PS cells.  The electric field of a $100\ \%$ polarized laser beam can be written as
\begin{equation}  \overrightarrow{E}=E_0(\cos\theta\cdot\hat{x}+e^{i\phi}\sin\theta\cdot\hat{y})\cdot e^{i(kz-\omega t)} = \left[\begin{array}{cc}\ \cos\theta \\ e^{i\phi}\sin\theta \end{array}\right],
\label{eq:elecvec}
 \end{equation}
where the vector expression is the Jones matrix notation \cite{optics}.  A useful method for characterizing the polarization of the beam utilizes the Stokes parameters.  If we assume that the beam is totally polarized and normalize the Stokes parameters to the intensity we then have
\begin{equation}
\begin{split}
S_1 &= \cos 2\theta = (X-Y)/(X+Y), \\
S_2 &= -\sin 2\theta \cos \phi = (U-V)/(U+V), \\
S_3^H &= -\sin 2\theta \sin \phi = (R^H-L^H)/(R^H+L^H), \\
(S_1^2 &+ S_2^2) + (S_3^H)^2 = L^2 + C^2 = 1,
\end{split}
\label{eq:stokes}
\end{equation}
where $S_1$ is a measure of linear polarization along the horizontal and vertical axes, $S_2$ is a measure of linear polarization along the axes at $\pm 45^{\circ}$ to the vertical, $S_3^H$ is a measure of the degree of circular polarization, X and Y represent intensities projected along the horizontal and vertical axes, U and V represent intensities projected along the axes at $\pm 45^{\circ}$ to the vertical, and $R^H$ and $L^H$ represent intensities projected onto a decomposition into right- and left-helicity circularly polarized light.  The superscript ``$H$'' is to indicate that we refer to the circular polarization in terms of its helicity for consistency with the particle physics definition, and $S_3^H$ is defined such that it is +1 for right-helicity circular polarization.  Two parameters, $S_1$ and $S_2$, are required to completely describe the linear polarization state of the beam.  We see from the last line of equations~\ref{eq:stokes} that the linear and circular polarization components are constrained to add in quadrature to a value of one.  One implication is that for a reasonably well circularly polarized beam, a small phase shift may have a negligible effect on the magnitude of the circular polarization while simultaneously inducing a large linear polarization.  For instance, a perfectly circularly polarized beam at $805\ \text{nm}$ which acquires a $2\text{-nm}$ phase shift passing through a thin low-stress window will have a circular polarization of $99.988\ \%$ and  a linear polarization of $1.6\ \%$.

The SLAC polarized source optical system includes elements such as the CP and PS Pockels cells, a half-wave plate, and additional optical elements that may each possess a small amount of birefringence.  These components can be well approximated as different cases of a unitary Jones matrix for a rotated retardation plate \cite{optics}:
\begin{equation}
J_{RET} = \left[\begin{array}{cc} 

\cos^2 \gamma + e^{i\delta}\sin^2 \gamma & (1 - e^{i\delta})\sin \gamma \cos \gamma \\
(1 - e^{i\delta})\sin \gamma \cos \gamma & \sin^2 \gamma + e^{i\delta}\cos^2 \gamma \\
\end{array}\right].
\label{eq:genret}
\end{equation}
Here, $\gamma$ is the angle between the retarder's fast axis and the horizontal axis and $\delta$ is the retardation induced between the fast and slow axes.  

The Cleanup Polarizer is oriented to transmit horizontally linearly polarized light; thus the initial electric vector can be represented as
\begin{equation}  \overrightarrow{E_i}=\left[\begin{array}{cc} 1 \\ 0 \end{array}\right].
\label{eq:ei}
 \end{equation}
The CP cell has its fast axis at $45^{\circ}$ from the horizontal ($\gamma = \pi/4$), induces a retardation $\delta_{CP}$, and can be represented by the matrix
\begin{equation}
J_{CP} = \frac{1}{2}\left[\begin{array}{cc} 

1 + e^{i\delta_{CP}} & 1 - e^{i\delta_{CP}} \\
1 - e^{i\delta_{CP}} & 1 + e^{i\delta_{CP}} \\

\end{array}\right].
\label{eq:cpdefn}
\end{equation}
The PS cell has its fast axis vertical ($\gamma = \pi/2$), induces a retardation $\delta_{PS}$, and can be represented by the matrix
\begin{equation}
J_{PS} = \left[\begin{array}{cc} 

e^{i\delta_{PS}} & 0 \\
0 & 1 \\

\end{array}\right].
\label{eq:psdefn}
\end{equation}

Calculating the state of the polarization vector immediately following the PS cell and multiplying both components by an additional phase shift in order to write the vector in a convenient form yields
\begin{equation}
\overrightarrow{E_f} = J_{PS} \cdot J_{CP} \cdot \overrightarrow{E_i} = \left[\begin{array}{cc} \cos \delta_{CP}/2 \\ e^{-i(\frac{\pi}{2}+\delta_{PS})} \sin \delta_{CP}/2 \end{array}\right].
\label{eq:ef}
\end{equation}
This optical configuration allows the generation of arbitrary elliptically polarized light.  Comparing equations~\ref{eq:elecvec} and~\ref{eq:ef}, we see that $\delta_{CP}$ determines the relative amplitude of the $x$ and $y$ components of the electric field, and $\delta_{PS}$ determines their relative phase.  Writing down the Stokes parameters for the light leaving the PS cell, we have
\begin{equation}
\begin{split}
S_1 &= \cos \delta_{CP}, \\
S_2 &= -\sin \delta_{CP} \sin \delta_{PS}, \\
S_3^H &= -\sin \delta_{CP} \cos \delta_{PS}. \\
\end{split}
\label{eq:stokesafter}
\end{equation}
As we indicated earlier, $\delta_{CP}$ is set to values close to $\pm\pi/2$ and $\delta_{PS}$ is set to values close to 0.  We see that for these values, the Stokes parameter $S_1$ is sensitive to small changes in $\delta_{CP}$, while $S_2$ is sensitive to small changes in $\delta_{PS}$.  Utilizing the Cleanup Polarizer and CP and PS cells in this configuration allows us to generate a laser beam of arbitrary elliptical polarization.  A convenient feature of this configuration is that any residual linear polarization can be decomposed into components that are separately adjustable by the CP cell ($S_1$) and the PS cell ($S_2$).

\subsubsection{Pockels Cell Alignment}  It is important to be able to properly align the Pockels cells with respect to the laser beam and to choose the portion of the crystal through which the beam passes.  To this end, we choose mounts for the Pockels cells that are adjustable in pitch, yaw, and roll and allow translation along both axes perpendicular to the beam.  The Pockels cells are initially aligned for pitch and yaw between crossed polarizers (the Cleanup Polarizer and an auxiliary analyzer), first adding the CP cell and recovering extinction, and then adding the PS cell.  Care is taken to be sure that the Pockels cell orientations are not at secondary minima.  The orientation of the Pockels cell fast and slow axes can be determined by then pulsing them one at a time at a high voltage ($\sim 2\ \text{kV}$) and adjusting the roll angle until extinction is recovered.  In this configuration, either the fast or slow axis is now parallel to the upstream polarizer.  The CP cell is then rotated by 45\textdegree; its orientation is verified and set more precisely later.  The PS cell is left in this orientation.

At this point, the analyzer is removed and the Helicity Filter\footnote{Meadowlark Optics, Frederick, Colorado, USA.} (HF) is used to check the alignment and set the initial Pockels cell voltages.  The HF consists of a linear polarizer and a quarter-wave plate fixed in orientation so that it transmits right-helicity light and extinguishes left-helicity light.  The nominal quarter-wave voltages for each helicity are set by sweeping the CP cell through the range $\pm(1500-3900\ \text{V})$ in 11 steps, measuring the transmitted light intensity, and fitting a parabola to the results.  Similarly, the nominal PS voltages are determined by setting the CP quarter-wave voltage for each state and sweeping the PS cell from $-1500\ \text{V\ to\ }+1500\ \text{V}$ in 11 steps, measuring the transmitted light intensity, and again fitting a parabola to the results.  To be satisfied with the alignment, we require that the extinction ratio between transmitted and extinguished states be greater than $\text{1000:1}$, that the sum of the CP right- and left-helicity voltages be below $100\ \text{V}$, and that the difference of the PS right- and left-helicity voltages be below $100\ \text{V}$.  The difference between the PS voltages is very sensitive to the alignment of the CP cell roll angle and provides the best means of verifying that it is properly oriented.  If the PS cell voltages are greater than $100\ \text{V}$ apart, the CP and PS cells are set to their left-helicity voltages so that they are extinguished by the HF, and the roll angle of the CP cell is adjusted to minimize transmission.  Then the voltage scans are repeated.  Requiring an extinction ratio of $\geq\text{1000:1}$ implies a circular polarization of $\geq99.8\ \%$ and an unpolarized component of $\leq0.2\ \%$.  An additional check of the voltages for right-helicity light, which are measured above in transmission, can be made by using the insertable half-wave plate mounted just upstream of the HF.  This allows us to measure the voltages for right-helicity light in extinction.  Finally, we check the quality of the laser beam polarization on the photocathode.  We do this by letting the beam strike the cathode and measuring the intensity asymmetry as the Pockels cell voltages are varied from their nominal values.  This procedure is described in more detail in section~\ref{strat_pita} and allows us to adjust the CP and PS cell voltages to compensate for residual birefringence in the optics between them and the photocathode.  A final cross check on the laser beam polarization is a scan of the Pockels cell voltages while measuring the electron beam polarization.  The only available electron beam polarimeter is the M\o ller polarimeter in End Station A, so this check can only be conducted while $\text{E-158}$ is running.  Typical operating voltages for the CP and PS cells are given in Table~\ref{tab:volts}, section~\ref{strat_pita}.

\subsubsection{Insertable Half-Wave Plates}
\label{sec:ihwp1}
We have two insertable zeroth-order half-wave plates in the optics system following the polarization optics that can be used to introduce a slow reversal of the laser helicity.  This flips the definition of helicity relative to what the data acquisition system (DAQ) is expecting, thus reversing the sign of the physics asymmetry.  Such a reversal is very useful for suppressing certain classes of systematic errors, and is discussed in section~\ref{sec:rev}.  One half-wave plate is located on the Helicity Control Bench (Figure~\ref{fig:hc}), where it is also useful for setting the initial Pockels cell voltages with the HF.  The second half-wave plate is located on the Cathode Diagnostics Bench (Figure~\ref{fig:otsgun}) and is the last optical element before the vacuum window at the entrance to the polarized electron gun.  Either half-wave plate can be used to effect the slow reversal, but we choose to use the one on the Cathode Diagnostics Bench for reasons discussed in section~\ref{sec:hwp}.  

\subsection{Helicity Control and Data Acquisition}  
\label{sec:DAQ}
\begin{figure}
   \centering
   \begin{tabular}{c}
   \includegraphics{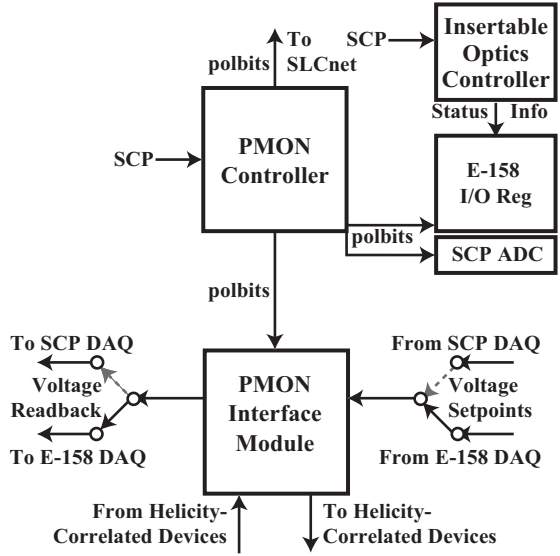}
   \end{tabular}
   \caption[example] 
   { \label{fig:pmon} 
A schematic of the Polarization MONitor (PMON) electronics.  Control of the helicity-correlated devices can be switched between the SCP and $\text{E-158}$ DAQs by swapping the ``Voltage Setpoints'' and ``Voltage Readback'' cables at the PMON Interface Module.  Additional copies of the polarization states (polbits) are available to an ADC read out by the SCP DAQ and to an $\text{E-158}$ I/O Register during commissioning but are eliminated during physics running.}
   \end{figure}

The beam helicity is controlled by the ``Polarization MONitor'' (PMON) system.  PMON's interaction with the optics hardware and the DAQs is shown in schematic form in Figure~\ref{fig:pmon}.  PMON is a set of SLAC-built custom electronics that generates a pseudorandom sequence of polarization states (``polbits'') using a $\text{33-bit}$ shift register algorithm as described in~\cite{Horowitz}.  At $120\ \text{Hz}$, this sequence repeats approximately once every two years.  Because the dominant noise in the electronic environment surrounding the accelerator is at $60\ \text{Hz}$, we choose to treat the $120\text{-Hz}$ triggering as two separate $60\text{-Hz}$ time slots.  We do this by imposing a quadruplet structure on the helicity sequence in which two consecutive pulses have randomly chosen helicities and the subsequent two pulses are chosen to be their complements.  For example, a possible sequence could be ``LRRL LLRR.''  In the data analysis, asymmetries are calculated for each pair of events, where pairs are formed between the first and third members of the quadruplet, and between the second and fourth members.  In this way we calculate pairwise asymmetries between pulses that are at the same phase with respect to the $60\text{-Hz}$ noise.  The pseudorandom sequence also provides a means of error checking in the offline analysis.  Observing the helicity state of 33 consecutive pairs allows one to predict the state of future pairs.  Comparing the predicted state with the actual state transmitted to the DAQ can be used to look for data acquisition errors.  PMON determines the pulse sequence, sets the appropriate voltages for all helicity-correlated devices (the CP, PS, and IA Pockels cells and the piezomirror), and distributes the helicity information and pulse identification number to the DAQs.

PMON interacts with two DAQs in order to control the helicity-correlated devices.  For testing and commissioning the source optics, we use PMON with the SLC Control Program (SCP).  For test beams and physics running, PMON works with the $\text{E-158}$ DAQ to control the optics.  Switching between the two DAQ systems is done by swapping a pair of cables at the PMON Interface Module that are used for the setting and readback of voltages for the helicity-correlated devices.  SCP is also used to control which algorithm the PMON Controller uses to generate the helicity sequence.  Five sequences are available:  the pseudorandom sequence described above, an alternating left/right sequence, all left-helicity pulses, all right-helicity pulses, and all no-helicity pulses (for which none of the helicity-correlated devices are operated).  The ``Insertable Optics Controller'' is a SLAC-built module which receives control signals from SCP and sends the appropriate voltage levels to the insertable optical elements.  It also sends status information to the $\text{E-158}$ I/O Register.

Three techniques are implemented in PMON to prevent the transmission of helicity information to the DAQs from introducing false asymmetries via electronic cross talk.  First, PMON delays the transmission of helicity information by one pulse.  This delay destroys the correlation between the actual beam helicity and the helicity information received by the DAQ.  Second, PMON converts the helicity information from a digital signal to an RF signal before transmitting it to the experiment over ``SLCnet,'' a dedicated  copper transmission line.  A PMON Receiver module in the \e158 DAQ decodes the RF signal.  Third, additional copies of the polarization information that are available as analog voltage levels during commissioning of the SCP and $\text{E-158}$ DAQs are eliminated for physics running.

\subsection{Laser Transport Optics and Cathode Diagnostics Bench}
\label{sec:ots}
We describe next the optical system between the polarization optics and the cathode.  These optics image the CP cell onto the cathode while preserving the circular polarization of the beam.  An ``Asymmetry Inverter'' consisting of two beam expanders with magnifications of equal magnitude and opposite sign (see Figure~\ref{fig:hc}) can be toggled between two positions to provide some cancellation for helicity correlations in the laser beam position and angle.  We used the software packages PARAXIA and ZEMAX to model gaussian beam propagation through the transport optics and to design the transport optics.

\subsubsection{Imaging}
\label{sec:imaging}
\begin{figure}
   \centering
   \begin{tabular}{c}
   \includegraphics{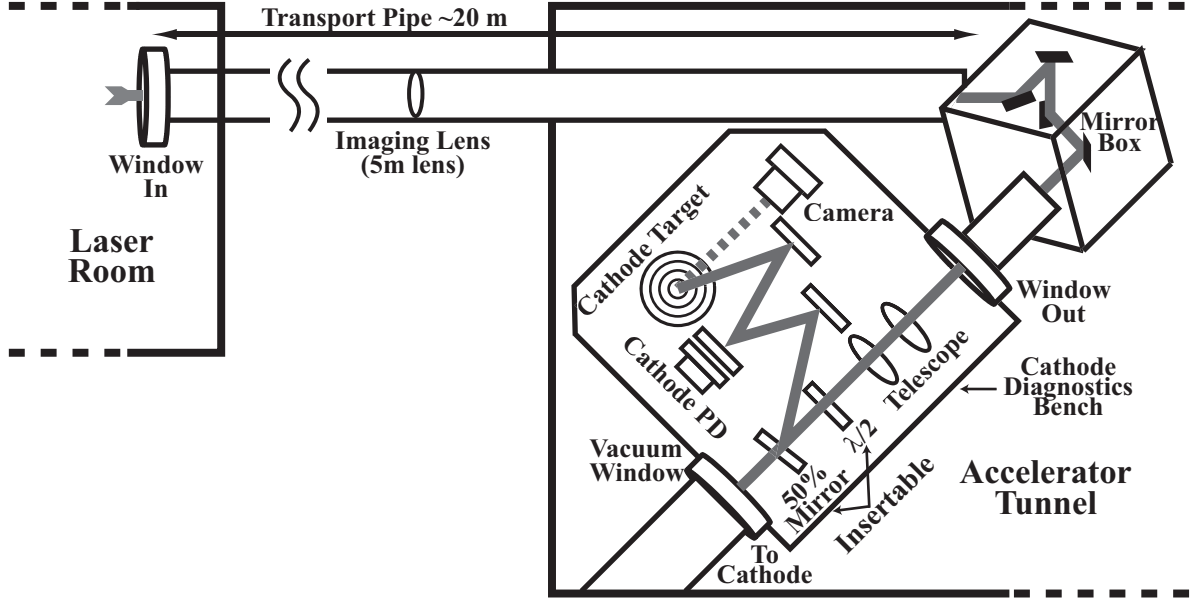}
   \end{tabular}
   \caption[example] 
   { \label{fig:otsgun} 
The optical transport system and the Cathode Diagnostics Bench.}
   \end{figure}

We image the CP cell onto the cathode in order to minimize the contribution of any helicity-correlated steering arising from the CP cell.  The imaging optics consist of the $5\text{-m}$ lens in the Transport Pipe and the telescope on the Cathode Diagnostics Bench (both shown in Figure~\ref{fig:otsgun}).  The location of the image point is most sensitive to the setting of the telescope on the Cathode Diagnostics Bench.  The downstream lens of that telescope is adjustable in $x$, $y$, and $z$, allowing us to set the beam size and position on the cathode and thereby dictating the location of the object point on the Helicity Control Bench.  We replaced the previous $3' \times 1'$ Helicity Control Bench with a $6' \times 16''$ bench to allow some freedom of movement to locate the CP cell at the object point.  The new cathode described in the introduction gives us additional flexibility to choose the laser spot size on the cathode.  Full illumination of the $20\text{-mm}$-diameter cathode is no longer needed to achieve the required electron beam current.  By reducing the spot size to $\sim 1\ \text{cm}$ for the $1/e^2$ diameter, we place the object point within a few centimeters of the CP cell and also improve the electron beam properties and transmission.  The imaging optics are designed to bring the laser beam through a waist between the telescope and the cathode to avoid clipping in the $14\text{-mm}$-diameter pipe that leads from the Cathode Diagnostics Bench to the cathode.  Because the laser beam gets as large as $1\ \text{inch}$ in diameter while being transported from the Laser Room to the cathode, we use $2\text{-inch}$ lenses and mirrors for the imaging optics and the Mirror Box to avoid clipping.

The remotely insertable $50\ \%$ mirror at the exit of the Cathode Laser Diagnostics Bench redirects the beam into a diagnostic line that has the same length as the distance between the mirror and the cathode. The Cathode Target provides an image of the beam as it appears on the cathode and is an extremely useful diagnostic, in particular for understanding the imaging of the beam and for measuring the position dependence of the cathode's QE.  We determine the location of the object point by placing a wire mesh screen in the beam near the CP cell and moving it along the beam axis while studying the quality of the image on the Cathode Target.  We observe that the object point is within a few centimeters of the CP cell.  This provides a significant reduction of the effective lever arm from the $\sim25\ \text{m}$ actual distance between the CP cell and the cathode. The object point is observed to be the same for both states of the Asymmetry Inverter.

\subsubsection{Asymmetry Inverter}
\label{sec:ai0}
The ``Asymmetry Inverter'' (AI) consists of two beam expanders on a translation stage as shown in Figure~\ref{fig:hc}.  The effect of the AI on an optics ray can be described by $\left| \widetilde{x} \right> = M\left| x \right>$, or equivalently
\begin{equation}
\label{eq:ai1}
\begin{bmatrix} \widetilde{x} \\ \widetilde{x}' \end{bmatrix} \quad = \quad
\begin{bmatrix} M_{11} & M_{12} \\ M_{21} & M_{22} \end{bmatrix} \begin{bmatrix} x \\ x' \end{bmatrix},
\end{equation}
where $x$ ($x'$) is the position (angle = $dx/dz$) of the optics ray entering the AI, $\widetilde{x}$ ($\widetilde{x}'$) is the position (angle) of the optics ray exiting the AI, and $M$ is the transport matrix characterizing the AI.  We designed the ``$+2.25I$'' and ``$-2.25I$'' optics to yield
\begin{equation}
\label{eq:ai2}
M^{\pm 2.25I} = \begin{bmatrix}  \pm 2.25 & 0 \\ 0 & \pm 0.44 \end{bmatrix}.
\end{equation}
Thus, the rays leaving the AI satisfy
\begin{equation}
\label{eq:ai4}
\left| \widetilde{x} \right>^{-2.25I} = -\left| \widetilde{x} \right>^{+2.25I}.
\end{equation}
The magnification of 2.25 is needed to assist in the beam transport to the cathode.  Taking equal amounts of physics data in the two AI configurations allows for cancellation of certain contributions to the $\alrb$'s and is discussed in section~\ref{ai}.  

\subsubsection{Preserving Circular Polarization}
\label{sec:preserve}
The transported laser beam must retain a high degree of circular polarization.  We employ several strategies to achieve this.  First, we minimize the number of optical elements in the transport system by placing the laser beam diagnostics in the auxiliary diagnostic line accessed by the insertable $50\ \%$ mirror.  Second, the four mirrors in the Mirror Box are arranged in two helicity-compensating pairs, for which the bounces within each pair interchange `s' and `p' polarizations.  Thus, the difference in phase shifts and losses between the `s' and `p' polarizations for a given mirror are cancelled between the members of each pair.  Care was taken to make certain that the four mirrors all came from the same coating run.\footnote{The mirrors are CVI TLM2-825-45-2037, specified to be from the same spindle and same coating run.  CVI Laser Corporation, Albuquerque, NM, USA.}  Finally, the Transport Pipe, which has historically been held under vacuum as part of a Class-IV laser containment system, is being used at atmospheric pressure for $\text{E-158}$ in order to minimize stress-induced birefringence in its end windows.  Alternate arrangements have been made to ensure the integrity of the Transport Pipe for laser safety purposes.

\subsection{Helicity-Correlated Feedbacks}
\label{sec:af}
Three active feedback loops are used to further suppress $\alrb$'s.  One feedback loop (the ``IA loop'') balances $A_I$ between the two helicity states.  This is accomplished by using the ``Intensity Asymmetry'' (IA) Pockels cell, located upstream of the Cleanup Polarizer (see Figure~\ref{fig:hc}).  When pulsed differently on right- and left-helicity pulses by a few tens of volts, the IA cell introduces a helicity-correlated phase shift into the beam.  The Cleanup Polarizer transforms this phase shift difference into an intensity asymmetry on the laser beam which compensates for the measured intensity asymmetry on the electron beam.  

A second feedback loop (the ``POS loop'') compensates for $D_{X(Y)}$ and $D_{X'(Y')}$.  This is accomplished by the ``piezomirror,'' a standard $1\text{-inch}$ diameter mirror attached to a Physik Instrumente model S-311.10\footnote{Physik Instrumente GmbH and Co, Auf der Roemerstrasse D-76228 Karlsruhe/Palmbach, Germany.} piezoelectric mount (see Figure~\ref{fig:hc}). This unit has three piezoelectric stacks that can be pulsed individually up to $100\ \text{V}$.  The independent operation of the three stacks gives the freedom to translate the face of the mount up to $6\ \mu\text{m}$, or to tilt in an arbitrary direction by up to $600\ \mu\text{rad}$.  The piezomirror changes the angle of the laser beam through the remainder of the optical system and can produce helicity-correlated displacements on the cathode of $20-50\ \mu\text{m}$ (depending on the effective lever arm between the piezomirror and the cathode), comparable to the beam position jitter.  

The third feedback loop (the ``Phase Feedback'') provides a mechanism for keeping the corrections induced by the IA loop small.  It looks at the correction induced by the IA loop averaged over a specified length of time and adjusts the CP and PS cell voltages in such a way as to drive the IA loop correction to zero.  Essentially, the Phase Feedback compensates for drifts in the polarization state of the laser beam that can give rise to an intensity asymmetry as described in section~\ref{sec:pita}.

The IA and POS loops utilize measurements from low-energy ($1\text{-GeV}$) electron beam diagnostics and act on the laser beam.  Additional independent diagnostics at both low and high ($45\ \text{GeV}$) energy are used to monitor the performance of the feedback loops and to measure $\alrb$'s at the $\text{E-158}$ target. We choose to generate the measurements for the feedback loops from beam diagnostics at low energy in order to minimize coupling between the various $\alrb$'s as a result of beam loading, residual dispersion, and wakefield effects in the accelerator.  The IA, POS, and Phase Feedback loops are discussed in more detail in section~\ref{sec:fdbk}.

%% file: n4.tex
\section[$\mathbf{Primary}$ $\mathbf{Sources}$ $\mathbf{of}$ $\mathbf{Helicity-Correlated\ Electron\ Beam}$ $\mathbf{Asymmetries}\ \mathbf{(^{beam}A_{LR}}\mathbf{'s)}$]{Primary Sources of Helicity-Correlated Electron Beam Asymmetries ($\alrb$'s)}
\setcounter{footnote}{0}
\label{sec:pita}
The primary mechanism for generating a helicity-correlated asymmetry in the intensity of the polarized electron beam, $A_I$, is a coupling between helicity-correlated changes in the orientation of residual linear polarization in the laser beam and the cathode's QE anisotropy.  The linear polarization components are a consequence of residual birefringence in the CP and PS cells and in the optics between them and the cathode.  This residual birefringence is significant:  a typical low-birefringence window produces a phase shift per unit thickness of $5\ \text{nm/cm}$.  The strained GaAs cathode's QE anisotropy provides a large analyzing power for incident linear polarization, typically on the order of $5-15\ \%$~\cite{Mair96}.  Uncorrected, a $5\text{-nm}$ phase shift can produce a helicity-correlated variation in electron beam intensity at the level of $0.4\ \%$, four orders of magnitude larger than the experimental requirement.  Similar polarization-related effects have sometimes been referred to~\cite{Cates} as PITA (Polarization-Induced Transport Asymmetry) effects and are often a dominant source of $\alrb$'s.  We derive an expression for $A_I$ for the case of the SLAC polarized electron source optics, identify the relevant phases, and examine the implications for controlling $\alrb$'s.  We also find that if the residual linear polarization of the laser beam varies spatially, it can give rise to helicity-correlated position and spot size differences.  We note that while in the analysis below the only analyzing power is a transport element in the optics system, the QE anisotropy of the cathode behaves formally in the same way.

\subsection{Derivation of the Polarization-Induced Transport Asymmetry}
\begin{figure}
   \centering
   \begin{tabular}{c}
   \includegraphics{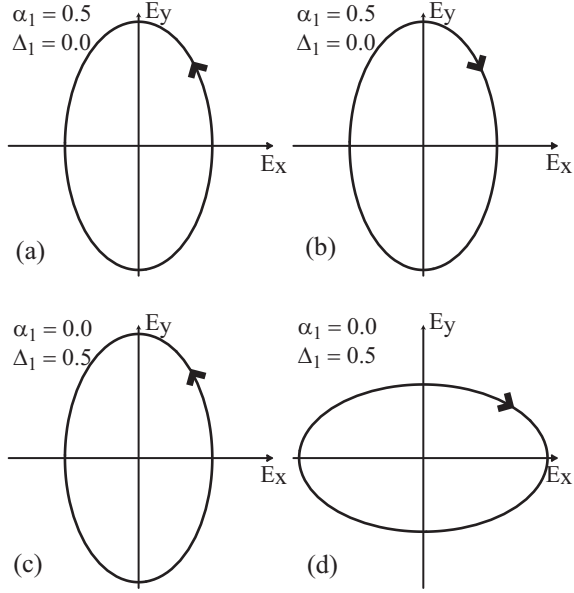}
   \end{tabular}
   \caption[example] 
   { \label{fig:ellipses} 
a) and b) Polarization ellipses generated for right- and left-helicity light, respectively, allowing $\alpha_1$ to be nonzero.  c) and d) Polarization ellipses generated for right- and left-helicity light, respectively, allowing $\Delta_1$ to be nonzero.  In each case, the other phases in equations~\ref{eq:phases} are set to zero.  The arrows indicate the direction of increasing time.}
   \end{figure}

We can understand the origin of $A_I$ by considering a system (as described in section~\ref{sec:wb}) comprised of horizontally polarized light incident in turn on the CP cell, the PS cell, and an asymmetric transport element.  The final electric vector can be computed by multiplying the initial electric vector by the appropriate matrices:
\begin{equation}  \overrightarrow{E_f}=J_{AT} \cdot J_{PS} \cdot J_{CP} \cdot \overrightarrow{E_i},
\end{equation}
where $J_{CP}$ and $J_{PS}$ are given by equations~\ref{eq:cpdefn} and~\ref{eq:psdefn} and $J_{AT}$ is an asymmetric transport element which provides an analyzing power that is sensitive to the orientation of linear polarization and does not introduce any depolarization.  Assuming the asymmetric transport element has transmission coefficients $T_{x'}$ and $T_{y'}$ along some axes $x'$ and $y'$, we have
\begin{equation}
J_{AT} = \left[\begin{array}{cc} 

T + \frac{\epsilon}{2}\cos 2\theta & \frac{\epsilon}{2}\sin 2\theta \\
\frac{\epsilon}{2}\sin 2\theta & T - \frac{\epsilon}{2}\cos 2\theta \\

\end{array}\right],
\label{eq:gendiatt}
\end{equation}
where $T = (T_{x'} + T_{y'})/2$, $\epsilon = T_{x'} - T_{y'}$, and $\theta$ is the angle between $x'$ and the horizontal axis.  The difference in transport efficiency along $x'$ and $y'$ is taken to be small ($\epsilon << T$).  Forming the intensity,
\begin{equation}
\begin{split}
I &= \overrightarrow{E_f^*} \cdot \overrightarrow{E_f} \\
&= T^2 + \frac{\epsilon^2}{4} + \epsilon T \cos \delta_{CP} \cos 2\theta - \epsilon T \sin \delta_{CP} \sin \delta_{PS} \sin 2\theta,
\end{split}
\label{eq:inten}
\end{equation}
we see that the final intensity of the beam is modulated by the phase shifts induced by the CP and PS cells and the orientation of the asymmetric transport element.  We allow the CP and PS cells to induce retardations that provide a fully general description of elliptically polarized light.  We choose a particular way to write them, however, so that the asymmetry has a simple form:
\begin{xalignat}{2}
\label{eq:phases}
\delta_{CP}^{R} &= -(\frac{\pi}{2}+\alpha_1)-\Delta_1, & \delta_{CP}^{L} &= +(\frac{\pi}{2}+\alpha_1)-\Delta_1,   \\
\delta_{PS}^{R} &= -\alpha_2-\Delta_2, & \delta_{PS}^{L} &= +\alpha_2-\Delta_2, \notag
\end{xalignat}
where the superscripts $R,L$ indicate right- and left-helicity light and the imperfect phase shifts have been parameterized in terms of ``symmetric'' ($\alpha$) and ``antisymmetric'' ($\Delta$) pieces such that $\alpha_1 = \Delta_1 = \alpha_2 = \Delta_2 = 0$ corresponds to perfectly circularly polarized light.  We give the phases from the CP cell a subscript ``1'' because Stokes parameter $S_1$ is particularly sensitive to $\delta_{CP}$.  Similarly, the phases induced by the PS cell carry a subscript ``2'' to emphasize that Stokes parameter $S_2$ is particularly sensitive to $\delta_{PS}$.  These sensitivities are evident from the Stokes vector for the light following the PS cell in this parameterization (where the small-angle approximation is made): 
\begin{xalignat}{2}
&S_1^R = -\alpha_1 - \Delta_1,& &S_1^L = -\alpha_1 + \Delta_1, \notag \\
\label{eq:StksCPPS}
&S_2^R = -\alpha_2 - \Delta_2,& &S_2^L = -\alpha_2 + \Delta_2, \\
&S_3^{H,R} = + 1,& &S_3^{H,L} = - 1. \notag
\end{xalignat}
The reason for the names ``symmetric'' and ``antisymmetric'' is apparent from Figure~\ref{fig:ellipses}.  A nonzero $\alpha$ phase shift (Figures~\ref{fig:ellipses}a and~\ref{fig:ellipses}b) turns circular polarization into elliptical polarization for which both helicities have the same major and minor axes, i.e., the phase shift affects the two polarization ellipses symmetrically.  A nonzero $\Delta$ phase (Figures~\ref{fig:ellipses}c and~\ref{fig:ellipses}d), however, results in elliptical polarization for which the two polarization ellipses have their major and minor axes interchanged, an antisymmetric behavior.

The parameterization given in equations~\ref{eq:phases} gives a completely general description of the elliptical polarization reaching the cathode, where additional phase shifts from components downstream of the CP and PS cells (providing they impose unitary transformations on the polarization vector) can be included as additional contributions to the $\alpha$'s and $\Delta$'s.  Two special cases, the addition of a slightly birefringent optic and the addition of an imperfect half-wave plate, are discussed below in section~\ref{sec:hwp}; in those cases we separate out from the $\alpha$'s and $\Delta$'s the contributions made by those optics in order to make explicit the helicity-correlated effects they induce.

Before calculating $A_I$, we can argue that to first order, only the antisymmetric phase shifts $\Delta_1$ and $\Delta_2$ contribute to it.  From equations~\ref{eq:StksCPPS}, we see that $S_1^R - S_1^L = -2\Delta_1$ and $S_2^R - S_2^L = -2\Delta_2$.  The helicity-correlated difference in the amount of linear polarization depends solely on the antisymmetric phases.  That this can give rise to $A_I$ can be seen by considering again the ellipses in Figure~\ref{fig:ellipses}.  If one imagines that the polarization ellipses are propagated through an asymmetric transport element with greater transmission along the vertical axis than the horizontal, it is clear that the ellipses with symmetric phase shifts are transmitted with equal intensity while the ellipses with antisymmetric phase shifts are not.

Finally, we insert equations~\ref{eq:phases} into equation~\ref{eq:inten} and calculate $A_I$.  We use the small-angle approximation and only keep terms that are first order in phase shifts and first order in $\epsilon$:
\begin{equation}
\boxed{A_{I} = \frac{I(\delta_{CP}^R,\delta_{PS}^R) - I(\delta_{CP}^L,\delta_{PS}^L)}{I(\delta_{CP}^R,\delta_{PS}^R) + I(\delta_{CP}^L,\delta_{PS}^L)} = -\frac{\epsilon}{T}[(\Delta_1 - \Delta_1^0) \cos 2\theta + (\Delta_2 - \Delta_2^0) \sin 2\theta].}
\label{eq:pita}
\end{equation}
We allow that residual birefringence in the Pockels cells or the optics downstream of them may introduce offsets by including the terms $\Delta_1^0$ and $\Delta_2^0$.  Note that birefringence in downstream optics can only contribute antisymmetric ($\Delta$-type) phase shifts.  The formalism above assumes that the asymmetric transport element is a component of the optical system.  Examples would include any optical element that is not exactly normal to the beam.  However, equation~\ref{eq:pita} remains valid if the optical analyzing power is replaced by a cathode with a QE anisotropy.  The strained GaAs cathodes in use at SLAC provide the dominant analyzing power in the system. 

\subsection{PITA Slopes}
\label{sec:slopes}
Note that $A_I$ depends linearly on the two antisymmetric phase shifts, $\Delta_1$ and $\Delta_2$.  This allows us to define two ``PITA slopes'' $m_1$ and $m_2$ that are easily measurable and characterize the sensitivity to residual linear polarization of a given optical system and analyzer.  The PITA slopes play a central role in our techniques for minimizing $\alrb$'s.  In practice, it is convenient to express the asymmetry formula in terms of these observables:
\begin{equation}
\begin{split}
m_{1} &= -\frac{\epsilon}{T}\cos 2\theta \\
m_{2} &= -\frac{\epsilon}{T}\sin 2\theta \\
A_{I} &= m_{1}\cdot(\Delta_1 - \Delta_1^0) + m_{2}\cdot(\Delta_2 - \Delta_2^0.)
\end{split}
\label{eq:slopes}
\end{equation}

\begin{figure}
   \centering
   \begin{tabular}{c}
   \includegraphics{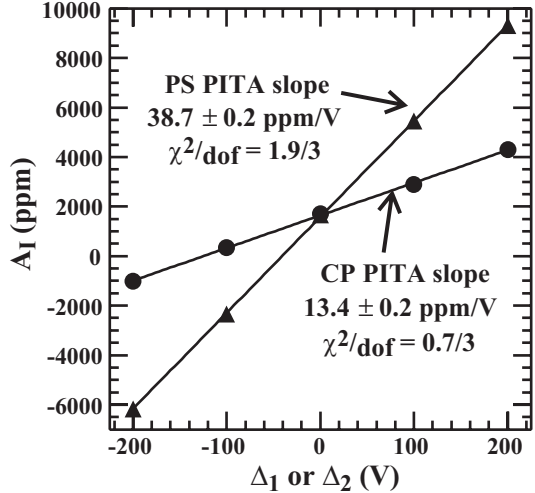}
   \end{tabular}
   \caption[example] 
   { \label{fig:pita} 
PITA slopes for both the CP and PS cells from \t437.  Slopes are plotted in terms of ppm of intensity asymmetry per volt of antisymmetric offset voltage applied to the Pockels cells and are typical values for strained GaAs cathodes.  Reprinted from~\cite{t437}.}
   \end{figure}

The phases $\Delta_1$ and $\Delta_2$ can be converted to voltages using equation~\ref{eq:mradvolts}.  By adjusting the voltages on the CP and PS cells in an antisymmetric fashion, one can adjust the size of either the Stokes 1 or the Stokes 2 components, and thus the size of $A_I$. For instance, suppose one optimizes the laser circular polarization after the PS cell (using the HF to maximize or minimize the transmitted light) and finds that the CP cell voltages should initially be set to $V_1^R = +2700\ \text{V}$ and $V_1^L = -2700\ \text{V}$.  To measure the CP cell's PITA slope, one applies offset voltages $\Delta_1$ and measures the resulting $A_I$ as shown in Figure~\ref{fig:pita}.  For $\Delta_1 = +200\ \text{V}$, we have $V_1^R = +2900\ \text{V}$ and $V_1^L = -2500\ \text{V}$ in this example.  Once the PITA slopes are measured and $A_I$ is measured for $\Delta_1 = \Delta_2 = 0$, offset voltages for $\Delta_1$ and $\Delta_2$ can be applied to null $A_I$.  This procedure is further described in section~\ref{strat}.

\subsection{Spatial Variation of Birefringence}
\label{sec:bigrad}
\begin{figure}
   \centering
   \begin{tabular}{c}
   \includegraphics{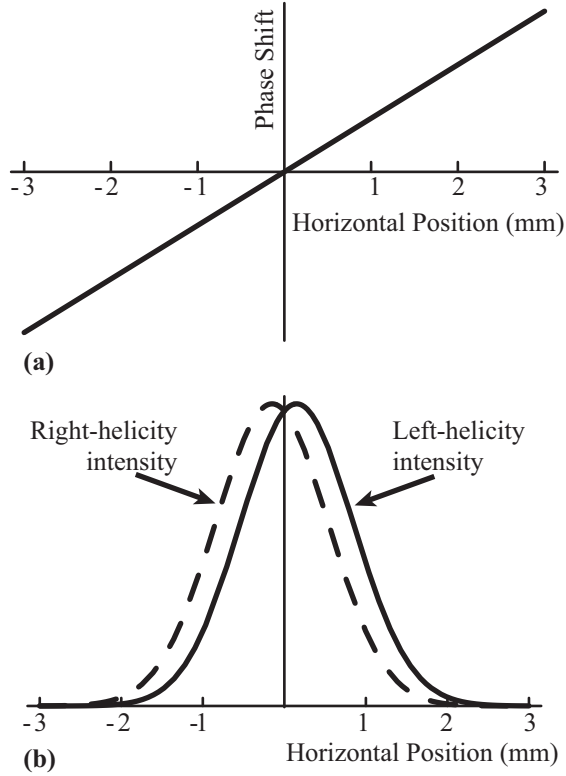}
   \end{tabular}
   \caption[example] 
   { \label{fig:bigradtheory}
 Demonstration of the effect of a linear gradient in the phase across the face of the laser beam on the resulting spatial intensity profile of the electron beam.  a)  The linear gradient in phase, exaggerated to demonstrate the effect.  b) The resulting intensity profiles for right- and left-helicity electron beams, assuming the incident laser beam had a gaussian profile.  The gradient shifts the centroids of the right- and left-helicity beams in opposite directions, generating a helicity-correlated position difference.
}
   \end{figure}

We have seen that $A_I$ is directly proportional to $\Delta_1 - \Delta^0_1$ and $\Delta_2 - \Delta^0_2$.  We have assumed, however, that if an optical element introduced a phase shift $\Delta$, the phase shift is the same regardless of the point on the face of the element through which the light passed.  But what if the phase shift varied across the face of the optical element?  If we allow that the residual birefringences $\Delta^0_1$ and $\Delta^0_2$ may have a spatial dependence to them, then it follows that $A_I$ will also have a spatial dependence.  A spatially varying $A_I$ opens the possibility of higher-order helicity correlations.  For instance, a laser beam with a $\Delta^0_1$ varying linearly in $x$ as in Figure~\ref{fig:bigradtheory}a produces an electron beam with a linearly varying $A_I$. This variation has the effect of shifting the centroids of the right- and left-helicity electron beams in opposite directions as illustrated in Figure~\ref{fig:bigradtheory}b and yields an electron beam with a helicity-correlated horizontal position difference.  Such effects are certainly present in the Pockels cells and are likely present at some level in the downstream optics as well.  

A spatially varying $\Delta$ retardation is one of the dominant sources of higher-order (position, spot size, and spot shape) helicity correlations in the spatial profile of the electron beam.  A convenient way to characterize the spatially varying phase shift is via ``moments'' similar to the moments of a statistical distribution.  Each moment can then be connected to a particular $\alrb$.  The zeroth moment (the average phase shift across the beam) gives rise to $A_I$.  The first moment is related to the gradient in phase shift across the beam and gives rise to $D_{X(Y)}$.  The second moment is related to the curvature of the phase shift across the beam and gives rise to spot size differences.  Similarly, higher-order moments can be related to higher-order helicity correlations in the beam profile.  Note from equation~\ref{eq:slopes} that the sensitivity of the electron beam to spatial variations in $\Delta^0_1$ and $\Delta^0_2$ scales with $m_1$ and $m_2$, respectively.  We discuss characterizing and minimizing such effects further in section~\ref{sec:ss}.  

\subsection{Vacuum Window Birefringence and Half-Wave Plate Cancellation}
\label{sec:hwp}
As we consider how to suppress $\alrb$'s arising from residual linear polarization, it is useful to separate out of the offset terms $\Delta_1^0$ and $\Delta_2^0$ the contributions that arise from the insertable half-wave plate used for slow helicity reversal and the vacuum window at the entrance to the polarized electron gun.  The vacuum window possesses a significant stress-induced birefringence and is unavoidably downstream of the half-wave plate and the Asymmetry Inverter, making it difficult to arrange cancellations of helicity-correlated asymmetries that arise from it.  For the remainder of the paper, we redefine $\Delta_1^0$ and $\Delta_2^0$ to exclude the residual birefringence associated with the vacuum window and the insertable half-wave plate, each of which we consider separately.  We model the vacuum window as a retardation plate with a small retardation $\beta$ and an arbitrary orientation angle $\rho$ measured from the horizontal axis.  The vacuum window can then be represented as 
\begin{equation}
J_{VW} = \left[\begin{array}{cc} 
\cos^2 \rho + e^{i\beta}\sin^2 \rho & (1 - e^{i\beta})\sin \rho \cos \rho \\
(1 - e^{i\beta})\sin \rho \cos \rho & \sin^2 \rho + e^{i\beta}\cos^2 \rho \\
\end{array}\right].
\end{equation}
The final electric field vector including the vacuum window is
\begin{equation}  \overrightarrow{E_f}=J_{AT} \cdot J_{VW} \cdot J_{PS} \cdot J_{CP} \cdot \overrightarrow{E_i}.
\end{equation}
The vacuum window contribution, having been separated out of $\Delta_1^0$ and $\Delta_2^0$, manifests itself as a third term in the asymmetry equation:
\begin{equation}
A_{I} = -\frac{\epsilon}{T}[(\Delta_1 - \Delta^0_1)\cos 2\theta + (\Delta_2 - \Delta^0_2)\sin 2\theta + \beta \sin (2\theta - 2\rho)].
\label{eq:bi}
\end{equation}

Next we consider the insertion of the half-wave plate used for slow helicity reversal, as discussed in sections~\ref{sec:ihwp1} and~\ref{sec:ihwp2}.  Here we focus on how the half-wave plate manipulates residual linear polarization.  We want to understand to what degree a cancellation of position and spot size differences can be achieved if they arise from spatial variations in the residual birefringence of particular optical elements.  

We assume that downstream of the PS cell we have an imperfect half-wave plate followed by the vacuum window.  The half-wave plate is allowed an arbitrary orientation $\psi$ and a deviation $\gamma$ from perfect half-wave retardation and can be represented as 
\begin{equation}
\label{eq:ihwp}
J_{HW} = \left[\begin{array}{ccc} \cos^2 \psi + e^{i(\pi + \gamma)}\sin^2 \psi & & (1 - e^{i(\pi + \gamma)})\sin \psi \cos \psi  \\ (1 - e^{i(\pi + \gamma)})\sin \psi \cos \psi & & e^{i(\pi + \gamma)}\cos^2 \psi + \sin^2 \psi \end{array}\right].
\end{equation}
The final electric vector is calculated as
\begin{equation}  \overrightarrow{E_f}=J_{AT} \cdot J_{VW} \cdot J_{HW} \cdot J_{PS} \cdot J_{CP} \cdot \overrightarrow{E_i},
\end{equation}
and the resulting intensity asymmetry is
\begin{equation}
\begin{split}
A_{I} = -\frac{\epsilon}{T}[&(\Delta_1 - \Delta_1^0) \cos (2\theta - 4\psi) - (\Delta_2 - \Delta_2^0) \sin (2\theta - 4\psi) \\
&- \beta \sin (2\theta - 2\rho) - \gamma \sin (2\theta - 2\psi) ].
\label{eq:hw}
\end{split}
\end{equation}
We compare this result to equation~\ref{eq:bi}, bearing in mind that the coefficients multiplying each term ($\Delta_1^0$, $\Delta_2^0$, $\beta$, and $\gamma$) may have a spatial dependence that could give rise to helicity-correlated position or spot size differences.  Comparing the first two terms of each equation, which include the contributions of all optics upstream of the half-wave plate, we see that they have acquired both a relative minus sign and a dependence on the orientation of the half-wave plate.  We gain some freedom to choose the PITA slopes by appropriately orienting the half-wave plate, but their values cannot be chosen independently.  The optimal cancellation of position and spot size differences would be gained by inserting the half-wave plate in an orientation such that the PITA slopes are unchanged, but the relative minus sign prevents that.  However, if one can arrange for one PITA slope to be much larger in magnitude than the other, then one can orient the half-wave plate to preserve the large PITA slope and perhaps still achieve a reasonable cancellation of effects arising from the upstream optics.  Unfortunately, this procedure requires control over the orientation of the cathode's analyzing power and such control is impractical.

Comparing the third terms of equations~\ref{eq:bi} and~\ref{eq:hw}, we see that the vacuum window contribution flips sign with insertion of the half-wave plate.  The sign flip prevents any cancellation of $\alrb$'s arising from optics downstream of the half-wave plate, motivating us to place the half-wave plate as far downstream as possible.

The half-wave plate itself introduces a fourth term that is proportional to the deviation of its retardation from $\pi$.  To the extent that this term is significant, it poses an obvious problem for arranging a cancellation.  

In summary, higher-order effects are not preserved but change in a complex way with insertion of the half-wave plate, resulting in a decrease in the amount of cancellation.  One further strategy that can be pursued is to measure $\alrb$'s for a number of half-wave plate orientations and empirically determine which provides the best cancellation, but again this is not feasible because the half-wave plate is not readily accessible.  None of these complications pose a problem for minimizing $A_I$, however:  one simply measures the new PITA slopes and adjusts $\Delta_1$ and $\Delta_2$ accordingly as is discussed more later.  

A second option for using a half-wave plate insertion to generate a slow helicity reversal is to insert the half-wave plate between the Clean-up Polarizer and the CP cell.  By inserting the half-wave plate with its fast axis at 45\textdegree\ to the horizontal, the initial linear polarization is rotated from horizontal to vertical and the sense of the circular polarization generated by the CP cell is reversed.  The half-wave plate matrix for this case is given by equation~\ref{eq:ihwp} with $\psi = 45$\textdegree.  The final electric vector is calculated as
\begin{equation}  \overrightarrow{E_f}=J_{AT} \cdot J_{VW} \cdot J_{PS} \cdot J_{CP} \cdot J_{HW} \cdot \overrightarrow{E_i},
\end{equation}
and $A_I$ becomes
\begin{equation}
\begin{split}
A_{I} = -\frac{\epsilon}{T}[&-(\Delta_1 - \Delta_1^0) \cos 2\theta - (\Delta_2 - \Delta_2^0)\sin 2\theta \\
&- \beta \sin(2\theta - 2\rho) + \gamma \cos 2\theta].
\end{split}
\end{equation}
The upstream half-wave plate inverts the sign of each of the first three terms relative to eqn.~\ref{eq:bi}.  Using the half-wave plate in this configuration is guaranteed to flip the sign of any $\alrb$'s arising from spatial variation in birefringence; no cancellation is gained.  We choose to use a half-wave plate placed as far downstream as possible, immediately before the vacuum window on the polarized gun, to gain the best cancellation possible, accepting that the cancellation is imperfect.

%% file: n5.tex
\section{$\mathbf{Techniques}$ $\mathbf{for}$ $\mathbf{Minimizing}$ $\mathbf{^{beam}A_{LR}}\mathbf{'s}$}
\renewcommand{\thefootnote}{\fnsymbol{footnote}}
\setcounter{footnote}{0}
\label{strat}
We have adopted a number of strategies for designing the polarized source optics system that are specifically aimed at minimizing $\alrb$'s.  Passive strategies include careful selection and setup of the CP and PS Pockels cells, imaging of the CP cell onto the cathode, and shaping of the laser pulse's temporal profile to compensate for beam loading effects.  Active strategies include feedbacks on $A_I$ and $D_{X(Y)}$.  Finally, introducing slow reversals of helicity correlations with respect to the physics asymmetry generates cancellations and provides a tool for studying systematic errors.  Each of these strategies is described in detail in the following subsections.  We finish this section by presenting results on the control of helicity-correlated asymmetries from \t437.

We are motivated to minimize $\alrb$'s via passive means as well as possible before using active feedbacks for two reasons.  First, an active feedback on a particular $\alrb$ can generate helicity correlations in other $\alrb$'s as a side effect.  Second, minimizing the $\alrb$'s we can control also likely suppresses higher-order $\alrb$'s that we cannot directly control and for which we have no active feedbacks.  Similarly, this concern motivates the use of ``Phase Feedback'' on the intensity asymmetry.  Another strategy is to monitor the higher-order moments of the electron beam's spatial intensity profile.  We are able to measure the beam's spatial intensity profile with a wire array located just upstream of the $\text{E-158}$ target.  These measurements allow us to determine whether higher-order $\alrb$'s are present in the electron beam at a significant level.

\subsection{Optimizing the CP and PS cells and the Laser Beam Polarization}
\label{pcs}
We optimize the CP and PS cells by selecting them for uniformity of retardation, orienting them relative to the beam carefully, setting their voltages to maximize the circular polarization after the PS cell, and then adjusting their voltages to minimize $A_I$ on the electron beam.  The basic setup of the polarization optics is discussed in section~\ref{sec:wb}.  Here we discuss those aspects of the setup that are specifically related to suppressing $\alrb$'s.

\subsubsection{Measuring the PITA Slopes and Correcting the Intensity Asymmetry}\label{strat_pita}  
\begin{figure}
   \centering
   \begin{tabular}{c}
   \includegraphics{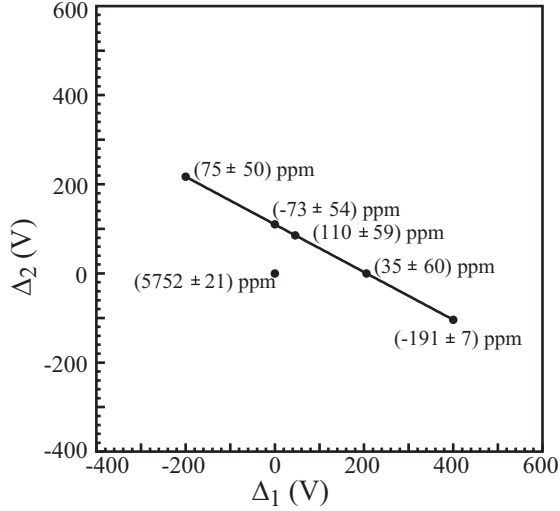}
   \end{tabular}
   \caption[example] 
   { \label{fig:space} 
Electron beam intensity asymmetry measured along the $A_I=0$ line and at $\Delta_1 = \Delta_2 = 0$.}
   \end{figure}

As was discussed in section~\ref{sec:slopes}, the PITA slopes characterize the sensitivity of a particular optical system and cathode to the presence of small linear polarization components and provide a key tool for minimizing $A_I$.  We determine the PITA slope for the CP cell by first measuring $A_I$ on the electron beam as we vary $\Delta_1$ in five steps over the range $\pm200\ \rm{V}$ (or $\sim\pm7\ \text{\textdegree}$, see equation~\ref{eq:mradvolts}) and then fitting a line to the resulting data.  Similarly varying $\Delta_2$ yields the PS cell PITA slope.  The PITA slopes for $\text{T-437}$ are shown in Figure~\ref{fig:pita}.  We can then define a ``voltage space'' representing a two-dimensional plane whose $x$- and $y$-axes correspond to $\Delta_1$ and $\Delta_2$.  Considering again equation~\ref{eq:slopes}, we see that if we set the left-hand-side equal to zero, we define a line in this voltage space along which the intensity asymmetry is zero, the ``$A_I=0$ line.''  As was discussed in section~\ref{sec:polcon}, the voltages determined for the CP and PS cells using the Helicity Filter need to be adjusted to compensate for phase shifts in the optics downstream of them.  This adjustment is equivalent to moving onto the $A_I=0$ line.  The $A_I=0$ line can be clearly seen in Figure~\ref{fig:space}.  These data were taken in the Gun Test Laboratory (on a different cathode than those used for the \e158 runs), which reproduces the first several meters of the accelerator.  The point of large $A_I$ at the origin is the intensity asymmetry measured using the nominal voltages determined via the Helicity Filter.  We have to choose a particular point on the line as optimal, and our strategy is to move toward the $A_I=0$ line in a perpendicular fashion in order to change the laser beam polarization by the minimum amount necessary to zero $A_I$.  The required $\Delta_1$ and $\Delta_2$ are then determined by 
\begin{xalignat}{3}
\Delta_1 = -\frac{A_{I} \cdot m_1}{m_1^2 + m_2^2},&  & \Delta_2 = -\frac{A_{I} \cdot m_2}{m_1^2 + m_2^2}
\label{eq:perp}
\end{xalignat}
where $A_{I}$ is the measured intensity asymmetry that needs to be corrected.

\begin{figure}
   \centering
   \begin{tabular}{c}
   \includegraphics{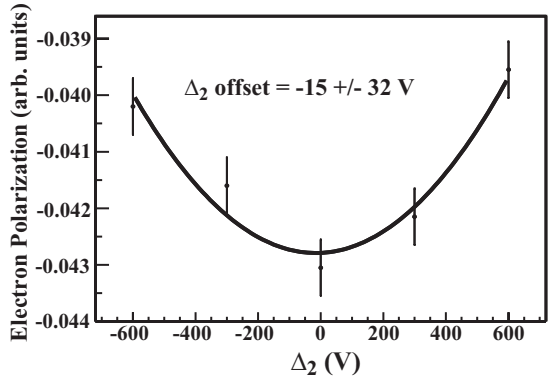}
   \end{tabular}
   \caption[example] 
   { \label{fig:moller} 
Electron beam polarization as a function of PS cell offset $\Delta_2$.  The CP cell offset $\Delta_1 = 0\ \text{V}$.  A parabola is fit to the data to approximate the $\cos(\frac{\pi \cdot \Delta_2}{V_{\lambda/2}})$ dependence for $\Delta_2 \ll V_{\lambda/2}$.}
   \end{figure}

We verify that this is a good choice of voltages by performing a similar scan of $\Delta_1$ and $\Delta_2$ and showing that the voltages which maximize the electron beam's polarization agree with those found by using equations~\ref{eq:perp}.  This check requires that the electron beam can be brought into End Station A in order to use the M\o ller polarimeter located there.  Figure~\ref{fig:moller} shows an example of such a study.  In Figure~\ref{fig:moller}, the electron beam polarization is measured as a function of $\Delta_2$, with $\Delta_1 = 0$.  The peak polarization is found to be at an offset voltage of $\Delta_2=(-15 \pm 32)\ \text{V}$, which is consistent with the values found by using equation~\ref{eq:perp} to move onto the $A_I=0$ line.  Table~\ref{tab:volts} summarizes typical operating voltages for the CP and PS cells as measured with the HF scans and after either using the PITA slopes to null $A_I$ or using the M\o ller polarimeter to measure the peak electron beam polarization.  
\begin{table}[h]
\caption{Typical operating voltages for the CP and PS cells for production of right- and left-helicity light for $\text{E-158}$ 2002 Physics Run I.} 
\label{tab:volts}\begin{center}       
\begin{tabular}{|l|l|l|l|l|} 
\hline
\hline
\rule[-1ex]{0pt}{3.5ex} & CP Right & CP Left & PS Right & PS Left \\
\hline
\rule[-1ex]{0pt}{3.5ex} HF Scan & 2607 V & -2732 V & -5 V & -9 V \\
\hline
\rule[-1ex]{0pt}{3.5ex} $\lambda$/2 OUT Null IA & 2574 & -2765 & -5 & -9 \\
\hline
\rule[-1ex]{0pt}{3.5ex} $\lambda$/2 OUT Polarimeter & 2582 $\pm$ 40 & -2757 $\pm$ 40 & -20 $\pm$ 32 & -24 $\pm$ 32 \\
\hline
\rule[-1ex]{0pt}{3.5ex} $\lambda$/2 IN Null IA & 2736 & -2603 & -105 & -109 \\
\hline
\rule[-1ex]{0pt}{3.5ex} $\lambda$/2 IN Polarimeter & 2667 $\pm$ 39 & -2672 $\pm$ 39 & -159 $\pm$ 35 & -163 $\pm$ 35 \\
\hline
\hline
\end{tabular}
\end{center}
\end{table}

\begin{figure}
   \centering
   \begin{tabular}{c}
   \includegraphics{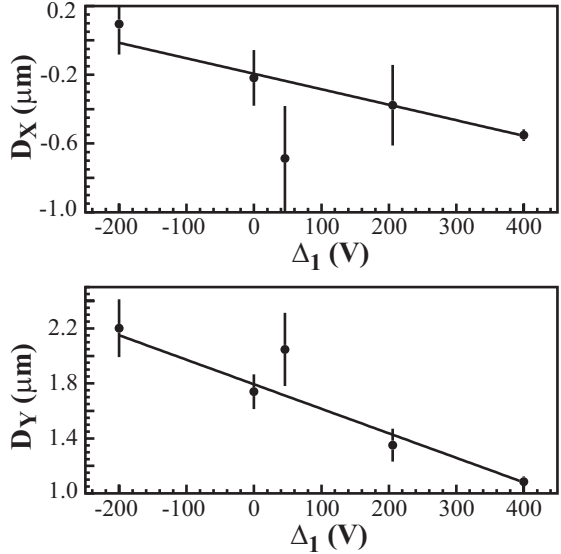}
   \end{tabular}
   \caption[example] 
   { \label{fig:line} 
Electron beam position differences as a function of $\Delta_1$ as $\Delta_1$ and $\Delta_2$ are simultaneously adjusted to slide along the $A_I=0$ line.}
   \end{figure}

It is also interesting to note that the position differences are typically sensitive to the choice of location along the $A_I=0$ line, as demonstrated in Figure~\ref{fig:line}.  These data were taken in the Gun Test Laboratory concurrent with the data shown in Figure~\ref{fig:space}.  Here, $D_X$ and $D_Y$ are plotted as a function of $\Delta_1$, with $\Delta_2$ correspondingly set to null $A_I$.  The physical mechanism underlying this behavior is not understood, although the sensitivity of position differences to the choice of $\Delta_1$ and $\Delta_2$ has been observed to depend on the ratio of $m_1$ to $m_2$ and on the choice of cathode.  One possibility is that this observation is caused by variations in the magnitude or orientation of the analyzing power across the face of the cathode.  In most cases, the dependence of $D_{X(Y)}$ on the choice of $\Delta_1$ and $\Delta_2$ was observed to be too small to provide a useful tool for minimizing $D_{X(Y)}$.  This fact also suggests that the mechanism which gives rise to this dependence is not one of the dominant mechanisms for generating $\alrb$'s in the SLAC system.

\subsubsection{Selection and Setup of Pockels Cells}  
\label{sec:ss}
\begin{figure}
   \centering
   \begin{tabular}{c}
   \includegraphics{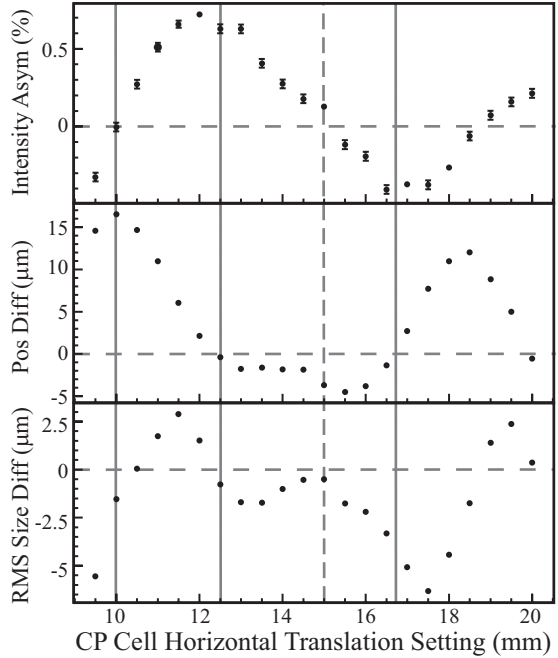}
   \end{tabular}
   \caption[example] 
   { \label{fig:bigrad} 
a) The helicity-correlated intensity asymmetry measured on the laser beam as a function of the Pockels cell's position as the Pockels cell is translated horizontally and the laser beam remains stationary.  b) Laser spot horizontal position difference vs. Pockels cell position.  c) Laser spot horizontal size difference vs. Pockels cell position.  The dashed vertical line indicates the position at which the laser beam is centered on the Pockels cell.  The solid vertical lines mark positions discussed in the text.}
   \end{figure}

The presence of spatially varying residual birefringence in the Pockels cells makes the selection of the Pockels cells and the setup of the beam through them important for suppressing helicity-correlated position and spot size differences.  We studied the residual birefringence in six QX2035 and one QX1020 Pockels cells and found that the peak-to-peak change in the residual birefringence across them varied between $\sim1.5-7\ \text{nm}$, depending on the Pockels cell.  A typical study of a Pockels cell, made in the Gun Test Laboratory, is shown in Figure~\ref{fig:bigrad}.  The measurements are made by placing a linear polarizer immediately after the Pockels cell in order to maximize the analyzing power.\footnote{In the case of a polarizer, the asymmetry expression must be modified because the polarizer does not satisfy the assumption $\epsilon \ll T$.  The asymmetry becomes $A_I = -\frac{4\epsilon T}{\epsilon^2+4T^2}[(\Delta_1 - \Delta_1^0) \cos 2\theta + (\Delta_2 - \Delta_2^0) \sin 2\theta]$.}  By orienting the polarizer at 45\textdegree\ to the Pockels cell's fast axis, we gain maximum sensitivity to variations in its residual birefringence.  The beam is detected by a linear array photodiode.\footnote{Model A2V-76, UDT Sensors Inc., Hawthorne, CA, USA.}  Twelve elements in the central portion of the array are instrumented, alternating instrumented and uninstrumented elements, providing a total detection area of $6.45\ \text{mm}\ \times\ 6.44\ \text{mm}$ with $50\ \%$ coverage.  The resulting signals are analyzed to determine the helicity-correlated intensity asymmetry and, along a single axis, the helicity-correlated position and spot size differences.  The position difference is obtained by computing the weighted mean of the position for each helicity according to $\overline{x}^{R(L)} = \sum_i (I_i^{R(L)}x_i^{R(L)}) / \sum_i I_i^{R(L)}$, where $x_i$ is the position of the $i^{\text{th}}$ element and $I_i^{R(L)}$ is the intensity measured by the $i^{\text{th}}$ element for right- (left-) helicity pulses.  The spot size difference is similarly calculated as the difference between the \emph{rms}'s for right- and left-helicity pulses.  The detector can be rotated by 90\textdegree\ to measure position and spot size differences along the other axis.  The detector is placed immediately after the polarizer in order to minimize the lever arm over which helicity-correlated lensing or steering differences can operate.  

For the particular study shown in Figure~\ref{fig:bigrad}, the polarizer was oriented to transmit vertically polarized light (yielding a PITA slope $m_1 = 550\ \text{ppm/V}$), the PS cell was removed, and the laser beam had a sigma of $1\ \text{mm}$.  The laser beam remained fixed in position while the CP cell was translated horizontally.
The three plots in Figure~\ref{fig:bigrad}, from top to bottom, show the intensity asymmetry, horizontal position difference, and horizontal size difference on the laser beam as a function of the horizontal position of the CP cell.  Based on the analysis given in section~\ref{sec:pita}, we interpret the variation in the intensity asymmetry as a spatial variation in the magnitude of the CP cell's residual birefringence.  The peak-to-peak variation in the intensity asymmetry of $1.1\ \%$ corresponds to a peak-to-peak variation in the phase shift of $1.5\ \text{nm}$, consistent with the expected level of residual birefringence variation for this model of Pockels cell.  There is a clear though imperfect correlation between the slope of the intensity asymmetry curve and the size of the position difference:  at a horizontal setting of 10, for instance, the intensity asymmetry reaches its maximal positive slope and the position difference also reaches an extremum.  Likewise, the position differences cross through zero at two points, 12.5 and 16.7, where the slope of the intensity asymmetry curve is near zero.  A similar imperfect correlation can be identified between the curvature of the intensity asymmetry curve and the size of the spot size difference.  For example, the spot size difference reaches extrema of opposite signs at horizontal settings of 11.5 and 17.5, points where the intensity asymmetry exhibits relatively large degrees of curvature of opposite signs.

The $\alrb$'s generated by birefringence gradients scale with the magnitude of the PITA slope; considering typical $18-50\ \text{ppm/V}$ PITA slopes arising from the analyzing power of the cathode and taking into account a factor of 2.5 magnification in spot size from the CP cell to the cathode, this Pockels cell would yield position differences as large as $1.3-3.8\ \mu\text{m}$ and spot size differences as large as $0.5-1.5\ \mu\text{m}$ on the electron beam.

We chose the two QX2035 Pockels cells with the smallest gradients in birefringence and least curvature to use as the CP and PS cells.  We placed them on two-axis translation stages in order to be able to optimize in situ the points on the crystals through which the laser beam passes.  Because the gradients across these two Pockels cells are significantly smaller than those observed on other cells and the cathode used during 2002 Physics Run I has an analyzing power that is a factor of two smaller than the previous cathode, the sensitivity of $\alrb$'s to their translation is significantly reduced.  In addition, the contributions to $\alrb$'s from these two Pockels cells are relatively small when they are centered on the beam (similar to the Pockels cell scan shown in Figure~\ref{fig:bigrad}).  We decided to run with them centered on the beam for 2002 Physics Run I.

We also observed that by minimizing the diameter of the laser beam at the Pockels cells we could further reduce our sensitivity to gradients in the residual birefringence.  However, because the Pockels cells are naturally birefringent,\footnote{The optic axis has $n_e = 1.4638$ and the transverse plane has $n_o = 1.5021$ at $\lambda=694\ \text{nm}$.} it is important to keep the beam well collimated passing through them for two reasons:  to ensure that all rays receive an equal retardation and to minimize the correlation between position and angle within the beam.  We designed the upstream optical transport system (consisting of the three lenses on the Flash:Ti, Diagnostics, and Helicity Control benches, see Figures~\ref{fig:ft} and~\ref{fig:hc}) to bring the beam through a gentle focus at the CP and PS cells.  Balancing the conflicting requirements that the beam be both small and well collimated, we found the optimum beam sigma to be approximately $1\ \text{mm}$ at the CP and PS cells.   

\subsection{Imaging and Transport Optics}
\label{sec:imaging2}
As was discussed earlier in section~\ref{sec:imaging}, we image the CP cell onto the cathode to minimize helicity-correlated lensing or steering differences that may arise from its high-voltage operation.  Studies of helicity correlations arising from the helicity-flipping Pockels cell were made in preparation for the Bates $^{12}$C experiment~\cite{kumar}. They clearly demonstrated that the Pockels cell produced helicity-correlated lensing or steering differences and that imaging could be used to suppress these differences.  While exhaustive studies of helicity-correlated lensing and steering differences have not yet been carried out on the SLAC system, we use a similar model of Pockels cell and have seen significant evidence for the presence of such effects in a variety of measurements.  Uncontrolled, we suspect that they have the potential to contribute to $\alrb$'s at a level comparable to or even greater than birefringence gradients.  We estimate that by imaging the CP cell to the cathode we have reduced the effective lever arm from $\sim25\ \text{m}$ to a few centimeters, thereby strongly suppressing any helicity-correlated lensing or steering differences that may be present.  We also optimized the transport optics to preserve a high degree of circular polarization as described in section~\ref{sec:preserve}.

\subsection{Beam Loading Compensation}
\label{sec:bl}
The primary mechanism for introducing a helicity-correlated energy asymmetry, $A_E$, into the electron beam is through ``beam loading.''  The electrons early in a pulse absorb power from an accelerating cavity as it accelerates them.  Electrons later in the pulse therefore find less power available to accelerate them.  If the electron pulse's temporal profile were flat, the later electrons would be accelerated less, creating an energy spread along the length of the pulse.  In addition, a further time dependence to the power available for acceleration is imposed by the time structure of the RF voltage applied to the cavities.  We compensate for this beam loading effect and the curvature of the applied RF voltage by shaping the laser pulse's intensity profile with TOPS as described in section~\ref{sec:TOPS} and Figure~\ref{fig:topsshape}.  The optimal temporal profile is sloped, with its intensity decreasing with time.  The average beam loading (over the length of the pulse) is $5\ \%$, and the energy spread is reduced to $0.1\ \%$ by the pulse shaping and other accelerator tuning techniques~\cite{decker}.  Beam loading introduces an anticorrelation between the intensity and the energy of the beam:  we would expect that the $0.5\ \%$ intensity jitter during 2002 Physics Run I should drive an energy jitter of $0.025\ \%$.  The energy jitter was typically measured to be $0.03\ \%$ during the run.  To the extent that the jitter in the beam energy is in fact driven by jitter in the beam intensity, $A_E$ tracks $A_I$ and benefits from the performance of the IA feedback loop described below.  If we achieve our goal of $A_I < 2 \cdot 10^{-7}$ then we should also achieve our goal of $A_E < 2 \cdot 10^{-8}$, providing other mechanisms do not contribute significantly.  

\subsection{Active Feedbacks on $\alrb$'s}
\label{sec:fdbk}
Active feedbacks on $A_I$ and $D_{X(Y)}$ provide additional suppression of these helicity-correlated asymmetries beyond what is achieved by the polarization optimization procedures and imaging.  The implementations of the IA, POS, and Phase Feedback loops are described in the following subsections.  The hardware for the feedback loops is described above in section~\ref{sec:af}.  First, we outline the general algorithm and discuss how the active nature of the feedback can suppress the mean value of the quantity being fed back on faster than one would expect from counting statistics.  Performance results for the feedback loops from \t437 are discussed in section~\ref{sec:perf}.

\subsubsection{Feedback Algorithm and 1/N Scaling}
The feedback loops each require three ingredients:  a control device capable of being driven in a helicity-correlated fashion, the optical system and cathode which generate $\alrb$'s, and a set of diagnostic devices on the electron beam to measure $\alrb$'s.  The measured beam asymmetry (or difference, in the case of position) can be represented as a sum of the contributions from each part of the loop,
\begin{equation}
A_{beam} = A_{ctrl} + A_{opt} + A_{stat},
\end{equation}
where $A_{ctrl}$ is the asymmetry induced by the control device, $A_{opt}$ is the helicity-correlated asymmetry caused by the optics and cathode, and $A_{stat}$ is the contribution from statistical jitter in the measured electron beam parameter.  The asymmetry is averaged over a ``minirun'' of M pulse pairs (where each pair consists of one pulse of each helicity) and then a correction is applied on the following minirun according to the general algorithm
\begin{equation}
\begin{split}
A_{ctrl}^1 &= 0, \\
A_{ctrl}^n &= A_{ctrl}^{n-1} - gA_{beam}^{n-1},
\end{split}
\end{equation}
where ``$g$'' is the gain of the loop and ``$n$'' is the number of the minirun.

An active feedback loop can cause the central value to converge to zero faster than one would naively expect based on a knowledge of the jitter in the measurement and the available statistics.  Consider an ideal feedback loop, in which the control device has perfect resolution, a unity gain is chosen, $A_{opt}$ is constant, and the statistical jitter in the beam parameter contributes a noise asymmetry $A_{stat}$ for each minirun.  The noise asymmetry contribution has a width $\sigma_{stat}$.  Then allowing the feedback to run for N miniruns, the algorithm outlined above generates the following behavior:
\begin{xalignat}{2}
A_{ctrl}^1 &= 0,& A_{beam}^1 &= A_{ctrl}^1 + A_{opt} + A_{stat}^1 \\
A_{ctrl}^2 &= A_{ctrl}^1 - A_{beam}^1,& A_{beam}^2 &= A_{ctrl}^2 + A_{opt} + A_{stat}^2, \notag \\
&= -A_{opt} -A_{stat}^1,& &= -A_{stat}^1 + A_{stat}^2, \notag \\
& \vdots& &\vdots \notag \\
A_{ctrl}^N &= A_{ctrl}^{N-1} - A_{beam}^{N-1},& A_{beam}^N &= A_{ctrl}^N + A_{opt} + A_{stat}^N, \notag \\
&= -A_{opt} -A_{stat}^{N-1},& &= -A_{stat}^{N-1} + A_{stat}^N. \notag 
\end{xalignat}
Averaging the $N$ miniruns then yields
\begin{align}
\overline{A} &= \frac{1}{N}\sum_{n=1}^N A_{beam}^n \\
&= \frac{1}{N}\left(A_{opt} + A_{stat}^1 - A_{stat}^1 + A_{stat}^2 - \dots -A_{stat}^{N-1} + A_{stat}^N\right) \notag \\
&= \frac{1}{N}\left(A_{opt} + A_{stat}^N\right) \notag
\end{align}
The active nature of the feedback arranges a cancellation of the contributions to the asymmetry arising from statistical jitter for all miniruns except the last.  We see that the mean measured asymmetry scales as $(A_{opt}+A_{stat}^N)/N$.  We refer to this as ``1/N scaling,'' in contrast with the normal statistical behavior in the absence of feedback for which the mean measured asymmetry scales as $A_{opt}+\sigma_{stat}/\sqrt{N}$.

\subsubsection{Intensity Asymmetry Feedback}
\label{sec:ia}
The IA loop is responsible for ensuring that $A_I$ converges to zero and that it does so rapidly enough to meet the requirements of equations~\ref{eq:lim} within the available statistics.  Because the intensity jitter is $\sim0.3-1.0\ \%$, there are not necessarily enough statistics ($\sim3 \cdot 10^8\ \text{pairs}$) in the experiment to ensure that the final asymmetry meets $\text{E-158}$'s requirement of $A_I < 2 \cdot 10^{-7}$, so we require that the IA loop provide sufficient 1/N scaling to achieve this requirement.

The IA Pockels cell controls the loop by introducing into the laser beam a helicity-correlated phase shift which the Cleanup Polarizer transforms into the desired intensity asymmetry correction.  The transmission through the Cleanup Polarizer is given by 
\begin{equation}
T = \cos^2\left(\frac{V - V_0^{IA}}{V_{\lambda/2}^{IA}}\cdot\frac{\pi}{2}\right),
\end{equation}
where $T$ is the transmission, $V_0^{IA}$ is an offset voltage which arises from residual birefringence or misalignment, and $V_{\lambda/2}^{IA}$ is the half-wave voltage of the IA cell.
  We reduce our sensitivity to drifts in $V_0^{IA}$ by running both states nominally at $99\ \%$ transmission.  Doing so also has the effect of reducing the amplitude of helicity-correlated voltage changes.  The feedback is initialized with both left and right states at the bias voltage $V_{B}$ that yields a bias transmission $T_B = 99\ \%$,
\begin{align}
A_{IA}^1 &= 0, \notag \\
V_R^1 &= V_B, \\
V_L^1 &= V_B. \notag
\end{align}
We apply a correction by increasing the attenuation on the appropriate state according to
\begin{align}
\label{eq:iaalgorithm}
A_{IA}^n &= A_{IA}^{n-1} - A_{I}^{n-1} \\
\text{IF}\ A_{IA}^n \leq 0\ \text{THEN}\quad V_L^n &= V_{\lambda/2}\cdot\frac{2}{\pi}\sin^{-1}\left[\left(1-T_B\cdot\frac{1-A_{IA}^n}{1+A_{IA}^n}\right)^{1/2}\right]+V_0, \notag \\
V_R^n &= V_B, \notag \\
\text{ELSE}\quad V_L^n &= V_B, \notag \\
V_R^n &= V_{\lambda/2}\cdot\frac{2}{\pi}\sin^{-1}\left[\left(1-T_B\cdot\frac{1+A_{IA}^n}{1-A_{IA}^n}\right)^{1/2}\right]+V_0. \notag 
\end{align}
The gain has been set equal to one.  The IA cell has a half-wave voltage of $\sim5800\ \text{V}$ and can be driven to $750\ \text{V}$, yielding a range of asymmetry correction on the order of $\pm3\ \%$, much larger than the typical induced correction of $(1-5)\cdot10^{-4}$.  

A number of factors limit the 1/N scaling we can achieve with the IA loop.  The ultimate limiting factor is the finite resolution of the beam current monitors, which in our case is typically $25\ \text{ppm/pair}$, negligible compared to the $\sim0.5\ \%$ intensity jitter.  The more severe limitation on 1/N scaling for our system comes from the fact that while we are running the feedback at $1\ \text{GeV}$, the important measurement of $A_I$ for the experiment is the one made in front of the $\text{E-158}$ target, two miles downstream.  Beam losses in the accelerator can add a new source of intensity jitter at a potentially significant level.  We are currently evaluating whether this source of jitter is small enough to allow adequate scaling of $A_I$ and are considering using a current monitor at $45\ \text{GeV}$ for the feedback.  In addition, matching the data set used in the offline analysis to that used by the feedback to generate 1/N scaling in the online analysis is a solvable but nontrivial challenge.  We should also note that to achieve full 1/N scaling one must run with a unity gain.

\subsubsection{Phase Feedback}
\label{sec:df}
As is discussed in section~\ref{sec:pita}, the dominant source of $A_I$ is the interaction between phase shifts in the optics and the cathode's analyzing power.  Phase Feedback refers to a second layer of feedback on $A_I$ which  averages over the IA loop correction for a number of miniruns (typically 30) and then adjusts the CP and PS cell voltages to null the IA loop correction.  This procedure provides compensation for drifts in the residual birefringence of the Pockels cells and the optics downstream of them that can cause the IA loop correction to wander at the $100\text{-ppm}$ level.  We adjust the CP and PS voltages by an amount proportional to the average asymmetry induced by the IA cell over the last N miniruns, $A_{ind}$, according to the prescription of equations~\ref{eq:perp} with $A_I$ replaced by $-A_{ind}$.  In principle, the Phase Feedback can be used alone to null $A_I$, but using two layers of feedback provides flexibility.  The IA loop can be run on a short time scale to take advantage of 1/N scaling, while the Phase Feedback can be run on a longer time scale that is appropriate for keeping the IA loop correction small.  The IA loop alone typically applies a correction at the $100\text{-ppm}$ level, which is three orders of magnitude larger than the physics asymmetry.  The Phase Feedback is capable of reducing the IA loop correction to the few ppm level when averaged over several days.  

\subsubsection{Position Difference Feedback}
\label{sec:pos}
$\text{E-158}$ requires that the helicity-correlated position differences $D_{X(Y)}$ be below 10 nm averaged over several months of data taking.  The position differences are typically several microns at the cathode when no effort is made to control them; with various optimizations (choice of Pockels cells, translation of Pockels cells, choice of operating voltages, and imaging) we achieve $D_{X(Y)}$ as small as $0.5-1.0\ \mu\text{m}$.  We also observe approximately an order of magnitude reduction in $D_{X(Y)}$ between the cathode and the $\text{E-158}$ target.  This is due in part to a drop in spot size from a $2.5\text{-mm}$ sigma at the cathode to $1\ \text{mm}$ at the target.  The remainder comes from emittance growth due to synchrotron radiation emission.  The emittance growth causes the spot size to increase, and the additional magnetic focusing required further suppresses the phase space available for  $D_{X(Y)}$.  The goal is to achieve 1-2 orders of magnitude of suppression of $D_{X(Y)}$ from an active feedback system.  The position jitter at the $\text{E-158}$ target is typically $\lesssim 50\ \mu\text{m}$ per pair so it should be possible to achieve a statistical error on $D_{X(Y)}$ of a few nm averaged over the entire run.  The statistical error bar will be small enough that additional suppression from 1/N scaling is not necessary.  We use the POS loop to suppress both $D_{X(Y)}$ and $D_{X'(Y')}$ simultaneously.  We can use one feedback on the laser beam to suppress two effects on the electron beam because they are not linearly independent.  There should only be position differences at the cathode; angle differences at the $\text{E-158}$ target arise because the cathode is not imaged onto it by the electron beam optics.

\begin{figure}
   \centering
   \begin{tabular}{c}
   \includegraphics{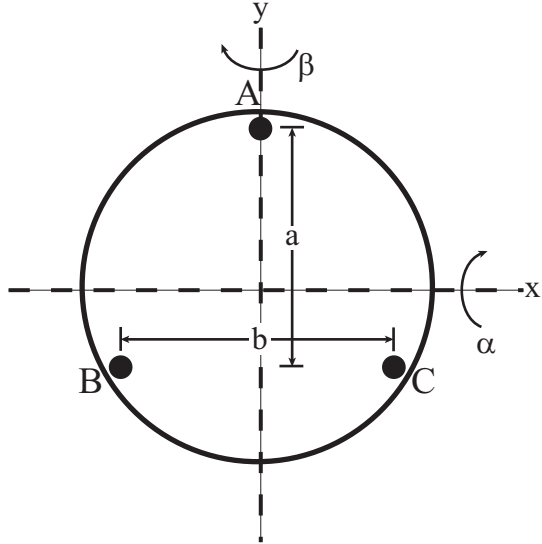}
   \end{tabular}
   \caption[example] 
   { \label{fig:pz} 
A schematic of the piezomirror.  The points labelled A, B, and C are the three piezo stacks.  The geometric factors a ($=10.4\ \text{mm}$) and b ($=12.0\ \text{mm}$) are relevant  for determining the rotation angles $\alpha$ and $\beta$ about the x- and y-axes, respectively.}
   \end{figure}

The piezomirror, described in section~\ref{sec:af}, is the control device for the POS loop.  A schematic of its operation is shown in Figure~\ref{fig:pz}.  The points A, B, and C represent the three piezo stacks.  The angles $\alpha$ and $\beta$ represent tilt about the $x$- and $y$-axes, respectively.  Pure tilts about the $x$- and $y$-axes can be achieved by applying voltages to A, B, and C according to 
\begin{align}
\label{eq:pzm}
\alpha &= \frac{A-\frac{1}{2}(B+C)}{a}, \notag \\
\beta &= \frac{B-C}{b}, \\
A &+ B + C = 0, \notag
\end{align}
where A, B, and C are given in millimeters of expansion.  The constraint that A, B, and C sum to zero implies that the piezomirror undergoes a pure tilt, with no translation.  When the feedback is initialized, it produces zero correction for the first minirun, and the voltages are set to
\begin{align}
\label{eq:pzinit}
&D^1_{Xpz} = 0\ \text{nm},\qquad D^1_{Ypz} = 0\ \text{nm}, \notag \\
&A^1_L = 0\ \text{V},\qquad B^1_L = 0\ \text{V},\qquad C^1_L = 0\ \text{V}, \\
&A^1_R = 0\ \text{V},\qquad B^1_R = 0\ \text{V},\qquad C^1_R = 0\ \text{V}, \notag 
\end{align}
where $D^1_{Xpz}$ and $D^1_{Ypz}$ are the position differences being induced by the POS loop on the first minirun and now $A^1_{R(L)}$, $B^1_{R(L)}$, and $C^1_{R(L)}$ are given in volts.  For subsequent miniruns, the induced position differences and piezomirror voltages evolve according to
\begin{align}
\label{eq:pzalgorithm}
&D^n_{Xpz} = D^{n-1}_{Xpz} - D^{n-1}_{X},\quad D^n_{Ypz} = D^{n-1}_{Ypz} - D^{n-1}_{Y}, \notag \\
&A^n_L = - A^n_R = \frac{a}{3l}\cdot D^n_{Ypz}\cdot 1.67\cdot 10^4 (\text{V/mm}), \\
&B^n_L = - B^n_R = \frac{1}{4l}\cdot (bD^n_{Xpz} - \frac{2}{3}aD^n_{Ypz})\cdot 1.67\cdot 10^4 (\text{V/mm}),  \notag \\
&C^n_L = - C^n_R = \frac{1}{4l}\cdot (-bD^n_{Xpz} - \frac{2}{3}aD^n_{Ypz})\cdot 1.67\cdot 10^4 (\text{V/mm}), \notag
\end{align}
where $D^{n-1}_{X}$ and $D^{n-1}_{Y}$ are the position differences measured on the $(n-1)^{th}$ minirun, $l$ is the effective lever arm from the piezomirror to the cathode, and the numerical factor is a conversion factor between voltage applied to a piezo stack and the resulting expansion.  

The piezomirror provided a range at the cathode of $\pm50\ \mu\text{m}$ helicity-correlated displacement (measured on the electron beam), and a range of $\pm4\ (\pm25)\ \mu\text{m}$ in $x$ ($y$) at the $1\text{-GeV}$ beam diagnostics during $\text{T-437}$.  The range has dropped to $\pm 20\ \mu\text{m}$ at the cathode and $\sim\pm(1-2)\ \mu\text{m}$ at $1\ \text{GeV}$ for the engineering and physics runs because of adjustment of the imaging optics and changes to the accelerator focusing lattice due to requirements for compatibility with other accelerator programs.  In fact, the constraints imposed on the electron optics by compatibility of multiple beams are severe enough that it was a challenge to make the POS loop work during 2002 Physics Run I.  The primary issue was that the phase advances of the electron beam in $x$ and $y$ from the cathode to the $1\text{-GeV}$ diagnostics differed by $\sim 90\text{\textdegree}$.  This made it difficult to find a tune for the electron optics that gave linearly independent beam motion at $1\ \text{GeV}$ for $x$- and $y$-tilts of the piezomirror.  We chose to run the POS loop with a gain of 0.25 to reduce our sensitivity to poor orthogonality between the $x$- and $y$-tilts.  We were free to choose the gain because we did not need 1/N scaling of the position differences.

\subsection{Slow Reversals and Cancellations of $\alrb$'s}
\label{sec:rev}
It is useful to incorporate into the experiment certain ``slow reversals'' that change the sign of an effect with respect to its influence on the physics asymmetry.  Such slow reversals serve two purposes:  they generate cancellations of $\alrb$'s and they provide a means of verifying our understanding of the measured detector asymmetry.  Having multiple slow reversals allows us to study the quality of cancellation each reversal provides.  If a single reversal yields the same central value for the physics asymmetry in both of its states then it provides a convincing cross check that false detector asymmetries (such as those arising from nonzero $\alrb$'s or electronic cross talk) are well understood.  However, if the two states disagree, it is difficult to quantify how to make a correction.  One can only average the results and hope that the false asymmetry contribution cancels out.  Using multiple reversals provides a number of complementary cross checks, each of which is sensitive to different combinations of false asymmetries.  Comparing the quality of agreement between states for each slow reversal may provide a measure of the uncertainty in the physics asymmetry.  

Two types of reversal can be identified:  those that are designed to reverse the sign of the physics asymmetry ($\alr^{\mo}$) without changing any false asymmetries in the experiment, and those that are designed to reverse the sign of certain $\alrb$'s without changing the physics asymmetry.  We have implemented one reversal of each kind in the source optics system:  the insertable half-wave plate provides a slow reversal of the physics asymmetry and the Asymmetry Inverter provides a slow reversal of position differences arising from the polarization optics.  These two reversals are discussed in more detail below.

Additional slow reversals can be implemented on the electron beam as well.  A second means of physics reversal, running the electron beam at two energies, takes advantage of the energy dependence of the rate of g-2 precession of the beam polarization as it is brought around a bend and into End Station A.  We run at two energies that each result in longitudinal polarization in End Station A but with a relative phase difference of 180\textdegree.  For a given electron helicity at the source, the two energies correspond to opposite helicities on target.  It would also be useful to implement a second Asymmetry Inverter in the electron beam optics, but this has not yet been done.

The slow and passive nature of these reversals is in marked contrast to the fast helicity reversal provided by the CP cell.  The CP cell provides a pulse-by-pulse helicity reversal, and, by virtue of its pseudorandom sequence, one that is insensitive to noise at low frequencies and to drifts in the experimental apparatus (such as gradual changes in the detector phototube gains).  The insertable half-wave plate and running at two energies both function as means of ``passively'' reversing the definition of helicity in the experiment in the sense that they do not involve electronic signals on a pulse-by-pulse basis, as the pulsing of the CP and PS cells do.  Certain classes of false asymmetries are not sensitive to the presence of the half-wave plate or the choice of electron beam energy.  These classes of false asymmetries cancel upon averaging data for the two states of each reversal.  One prime example is electronic cross-talk between the CP cell high voltage pulse and the data acquisition electronics.  

\subsubsection{Half-Wave Plate Reversal}
\label{sec:ihwp2}
The half-wave plate is expected to provide a good cancellation for many classes of false asymmetries, including those arising from electronic cross talk and helicity-correlated lensing or steering differences from the CP cell.  However, as is described in section~\ref{sec:hwp}, the behavior of $\alrb$'s arising from imperfections in the circular polarization of the laser beam depends on the orientation of the half-wave plate in a complicated way, so it is not straightforward to gain a cancellation of these effects.  We have two insertable half-wave plates, one on the Helicity Control Bench (which is primarily used for initially determining the operating voltages for the polarization Pockels cells) and one on the Cathode Diagnostics Bench.  For physics running we use the Cathode Diagnostics Bench half-wave plate because it is downstream of all other optics except the vacuum window on the electron gun.  We toggle the state of the half-wave plate once every 48 hours.  

\subsubsection{Asymmetry Inverter Reversal}
\label{ai}
Immediately downstream of the PS cell is the Asymmetry Inverter (AI), which is described in section~\ref{sec:ai0}.  Switching between its ``$+2.25I$'' and ``$-2.25I$'' optics inverts both the position and angle of the outgoing optics rays.  Averaging over all optics rays in the beam, any helicity-correlated position and angle differences resulting from the polarization optics should have opposite signs for the two AI states.  We can see this by considering the helicity-correlated position and angle differences for the laser beam entering the AI, averaging over all its optics rays:
\begin{equation}
\begin{split}
D_X &= \left< x \right>_R - \left< x \right>_L \\
D_{X'} &= \left< x' \right>_R - \left< x' \right>_L.
\end{split}
\end{equation}
Then for the laser beam exiting the AI, using equation~\ref{eq:ai4} we expect
\begin{equation}
\begin{split}
D_{\widetilde{X}}^{+2.25I} &= -D_{\widetilde{X}}^{-2.25I} = 2.25D_X \\
D_{\widetilde{X}'}^{+2.25I} &= -D_{\widetilde{X}'}^{-2.25I} = 0.44D_X'. 
\end{split}
\end{equation}
Alternate running with equal amounts of physics data in the ``$+2.25I$'' and ``$-2.25I$'' modes should yield
\begin{equation}
\begin{split}
D_{\widetilde{X}} &= D_{\widetilde{X}}^{+2.25I} + D_{\widetilde{X}}^{-2.25I} = 0 \\
D_{\widetilde{X}'} &= D_{\widetilde{X}'}^{+2.25I} + D_{\widetilde{X}'}^{-2.25I} = 0, 
\end{split}
\end{equation}
cancelling any position and angle differences which arise from the polarization optics.  However, the AI does not help with cancelling $\alrb$'s arising from downstream optics, and it does not allow cancellation of spot size differences due to their even symmetry.  

\subsection{Measurements of $\alrb$'s During \t437}
\label{sec:perf}
Measurements of $\alrb$'s were made during $\text{T-437}$ and are reported fully in~\cite{t437}.  $\text{T-437}$ was a one-week beam test in November 2000 which recommissioned the polarized electron source and commissioned the polarization optics, electron beam diagnostics, and SCP and \e158 DAQs that are used for controlling $\alrb$'s.  In particular, $\text{T-437}$ was an opportunity to commission the IA and POS loops, measure $A_I$ and $D_{X(Y)}$, and study the effectiveness of the half-wave plate and Asymmetry Inverter cancellations.  Here we summarize those results.

\subsubsection{Measurements of $A_I$}
\begin{figure}
   \centering
   \begin{tabular}{c}
   \includegraphics{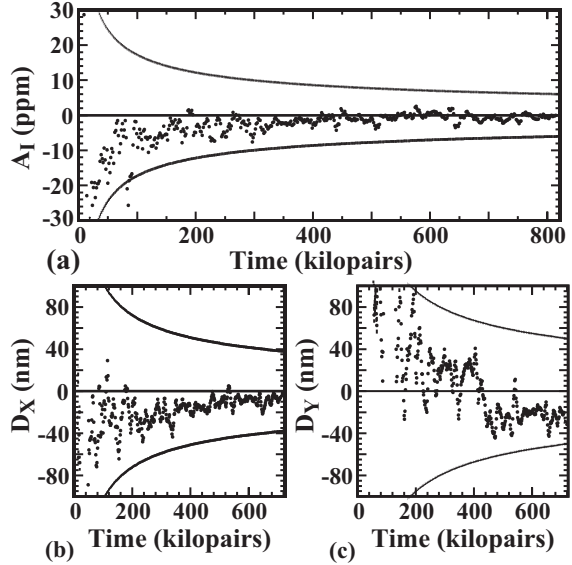}
   \end{tabular}
   \caption[example] 
   { \label{fig:intasyms} 
Average values of $A_I$, $D_X$, and $D_Y$ as a function of time during $\text{T-437}$.  Reprinted from~\cite{t437}.}
   \end{figure}

We commissioned the IA loop and measured $A_I$ during $\text{T-437}$.  A pair of toroids at the $1\text{-GeV}$ point in the accelerator measured the beam current and $A_I$.  The measurement from one toroid controlled the feedback while the second toroid provided an independent measurement that was used to evaluate the systematic error associated with the measurement and to determine the toroid resolution.  We used miniruns of $2000\ \text{pairs}$ in length, yielding a statistical error per minirun of $\sim5\cdot 10^{-4}$, based upon the $\sim0.6-0.8\ \%$ intensity jitter at that time.  Figure~\ref{fig:intasyms}a shows the performance of the IA loop during part of $\text{T-437}$.  The mean value of $A_I$ and an envelope representing the one-sigma statistical error (assuming a zero systematic offset) are plotted as a function of time in units of the number of pulse pairs included in the average.  For the entire $\text{T-437}$ test, we measured $A_I = 0.2 \pm 5.7\ \text{ppm}$, where the error bar is the statistical error (i.e., the rms width of the intensity asymmetry distribution divided by the square root of the number of pairs in the distribution).  The mean value of $0.2\ \text{ppm}$ is consistent with the expected asymmetry of $\sim0.33\ \text{ppm}$ assuming 1/N scaling.  Further evidence for 1/N scaling can be seen in Figure~\ref{fig:intasyms}a, where the mean value of $A_I$ rapidly acquires a value significantly less than the statistical error and wanders around zero.  $\text{T-437}$ also demonstrated  agreement between the two current monitors of $6 \pm 12\ \text{ppb}$, where the error is determined by the toroid resolution.  

\subsubsection{Measurements of $D_{X(Y)}$}
We also commissioned the POS loop and measured $D_{X(Y)}$ during $\text{T-437}$.  For that test, we used three beam position monitors (BPMs) spaced at $1\text{-m}$ intervals and located at the $1\text{-GeV}$ point in the accelerator.  The middle BPM was the control device for the feedback loop. The position differences measured by the upstream and downstream BPMs were used to predict the position difference at the middle BPM and thereby provide an independent measurement that was used to evaluate the systematic error associated with the measurement and to determine the BPM resolution.  We used a minirun length of $10,000\ \text{pulses}$. $\text{T-437}$ demonstrated that the POS loop closes, as shown in Figures~\ref{fig:intasyms}b and \ref{fig:intasyms}c.  $\text{T-437}$ set a limit at the $20\text{-nm}$ level on the ability of the POS loop to make position differences small and was only limited by the available statistics.  The corrections induced by the POS loop are tabulated in Table~\ref{tab:AI}.  The measurements at the middle (feedback) BPM were found to agree with the predicted value determined by the upstream and downstream BPMs to a level of $2\ \text{nm}$.

\subsubsection{Half-Wave Plate Cancellation}
We toggled the state of the half-wave plate on the Cathode Diagnostics Bench every two hours during several shifts of $\text{T-437}$ and found that it provided some cancellation of $D_{X(Y)}$, as shown in Table~\ref{tab:hw}.  For this running period, the IA loop was under control of the SCP DAQ and used a toroid in the injector for its feedback device; the POS loop was not used.  The cancellation yields approximately a factor of three reduction in $y$ and an order of magnitude in $x$.  However, the degree of cancellation is extremely sensitive to the details of the source setup, so this past performance cannot be taken as a strong indication of the quality of cancellation from the half-wave plate for future running.  A new cathode, realignment of the CP and PS cells, and adjustment of the imaging optics all can affect the ability of the half-wave plate toggling procedure to cancel $D_{X(Y)}$.  
\begin{table}[h]
\caption{Half-wave plate cancellation results.  The state of the half-wave plate was toggled every two hours over several shifts.  These results are the average measured position differences for that period of time.  In calculating the weighted average of the half-wave plate ``IN'' and ``OUT'' results, the sign of the ``OUT'' data is flipped to reflect the reversal of the sign of the physics asymmetry.  Reprinted from~\cite{t437}.} 
\label{tab:hw}
\begin{center}       
\begin{tabular}{|l|c|c|c|} 
\hline
\hline
\rule[-1ex]{0pt}{3.5ex} Asymmetry & $\lambda/2$ IN & $\lambda/2$ OUT & Weighted Average   \\
\hline
\rule[-1ex]{0pt}{3.5ex} BPM 2 X & 126 $\pm$ 12 nm & 119 $\pm$ 11 nm & -7 $\pm$ 8 nm   \\
\hline
\rule[-1ex]{0pt}{3.5ex} BPM 2 Y & -1732 $\pm$ 57 nm & -581 $\pm$ 51 nm & -461 $\pm$ 38 nm   \\
\hline
\hline
\end{tabular}
\end{center}
\end{table}

\subsubsection{Asymmetry Inverter Cancellation}
A test of the Asymmetry Inverter during T-437 yielded encouraging results.  Table~\ref{tab:AI} lists the corrections applied by the POS loop for runs in each state of the Asymmetry Inverter, along with their average.  For these runs, both the IA and POS loops were used under the control of the \e158 DAQ.  In the case of the $y$ position differences, for which the correction applied by the POS loop was large, a significant degree of cancellation is seen.  It should be noted that in both states the POS loop maintained the measured values of $D_{X(Y)}$ at a level that was consistent with zero.
\begin{table}[h]
\caption{Size of corrections applied by the POS loop for alternate states of the Asymmetry Inverter.  Reprinted from~\cite{t437}.} 
\label{tab:AI}
\begin{center}       
\begin{tabular}{|l|l|l|l|} 
\hline
\hline
\rule[-1ex]{0pt}{3.5ex} Fdbk Correction Induced & -2I State & +2I State & Average   \\
\hline
\rule[-1ex]{0pt}{3.5ex} $D_X$ Correction & 262 $\pm$ 48 nm & -39 $\pm$ 71 nm & 168 $\pm$ 40 nm   \\
\hline
\rule[-1ex]{0pt}{3.5ex} $D_Y$ Correction & -862 $\pm$ 109 nm & 1157 $\pm$ 130 nm & -29 $\pm$ 84 nm   \\
\hline
\hline
\end{tabular}
\end{center}
\end{table}

\subsection{Performance of Passive and Active Suppression Strategies}
We finish this section by comparing the performance of the passive optimization and the active feedbacks.  The intensity asymmetry is initially $0.1-0.5\ \%$ after setting the CP and PS cell voltages with the HF.  Adjusting $\Delta_1$ and $\Delta_2$ manually to null it brings it to the $100\text{-ppm}$ level.  The IA loop reduces $A_I$ by more than 2 orders of magnitude to $\ll 1\ \text{ppm}$ (for example, \t437 achieved $A_I \sim 0.3\ \text{ppm}$ in a short run and was statistics-limited).  The Phase Feedback can reduce the average IA loop correction to the few ppm level.

Helicity-correlated position differences are typically several microns prior to the passive optimization.  Optimizing the Pockels cell and laser beam setup and imaging the CP cell onto the cathode reduce $D_{X(Y)}$ to $0.5-1.0\ \mu\text{m}$.  The POS loop further reduces $D_{X(Y)}$ by $\sim 2$ orders of magnitude at the $1\text{-GeV}$ point in the accelerator (for example, \t437 achieved $D_{X(Y)} \sim 20\ \text{nm}$ and was statistics limited). 

%% file: n6.tex
\section{$\mathbf{Summary}$ $\mathbf{of}$ $\mathbf{Potential}$ $\mathbf{Sources}$ $\mathbf{of}$ $\mathbf{^{beam}A_{LR}}\mathbf{'s}$}
\setcounter{footnote}{0}
\label{sec:other}
In this section we review the mechanisms for generating $\alrb$'s which we have identified as dominant for the SLAC source.  We also mention several potential mechanisms which are not dominant but are possibly present at a smaller level.  Table~\ref{tab:table} provides a summary of the discussion given in the following subsections.  

The mechanisms that dominate for the SLAC source may not be the same as those that dominate for other sources.  Particular properties of the SLAC source that may influence which mechanisms are important include the particular model of Pockels cell in use (Cleveland Crystals QX2035), the particular cathode used (strained GaAs for \t437 and the 2001 engineering run; gradient-doped strained GaAsP for 2002 Physics Run I), the long physical distance between the polarization optics and the cathode ($\sim25\ \text{m}$), and the large diameter of the laser spot on the cathode ($1\ \text{cm}$ 1/e$^2$ diameter.).

\subsection{Residual Linear Polarization Coupling to QE Anisotropy}
The primary mechanism for generating $A_I$ is residual linear polarization interacting with asymmetric transport elements, as is discussed in section~\ref{sec:pita}.  The key optical elements that introduce residual linear polarization are the CP and PS Pockels cells and the vacuum window, each of which possesses significant residual birefringence.  Many optical components are candidates to be asymmetric transport elements, and care must be taken in setting up the optical transport system to minimize such effects.  The dominant analyzing power in the system is provided by the cathode's QE anisotropy.  $A_I$ can be suppressed either by reducing the amount of residual linear polarization incident on the cathode or by reducing the cathode's QE anisotropy.  Neither the half-wave plate nor the AI provide cancellations of this effect.  

\subsection{Birefringence Gradients}
As is discussed in sections~\ref{sec:bigrad} and~\ref{sec:ss}, gradients in the residual birefringence of the CP and PS cells and the vacuum window (and possibly other optics at a lower level) produce a spatial variation in $A_I$ and are a dominant source of helicity-correlated position and spot-size differences.  We suppress the contributions from the Pockels cells by carefully selecting and setting up the Pockels cells and by minimizing the beam diameter at their location.  The half-wave plate provides an imperfect cancellation (see section~\ref{sec:hwp}) of both position and spot size differences arising from these gradients for optics upstream of it and no cancellation for effects arising from the vacuum window.  The Asymmetry Inverter provides some cancellation of position differences from the Pockels cells.

\subsection{Lensing}
Helicity-correlated lensing or steering generated by the CP cell (sections~\ref{sec:imaging} and~\ref{sec:imaging2}) is present in the SLAC system at a significant level.  While it is difficult to untangle the relative contributions of lensing and birefringence gradient effects, we believe that imaging has suppressed lensing sufficiently that it is comparable to or smaller than the birefringence gradient effects.  When we have isolated effects that can be interpreted as lensing in laser beam measurements, we have observed that the effects cancel with the half-wave plate insertion.  In addition, we expect that the Asymmetry Inverter also provides a cancellation of position differences.

\subsection{Beam Loading}
As discussed in section~\ref{sec:bl}, beam loading is the primary mechanism for inducing an energy asymmetry.  We compensate for beam loading by properly shaping the laser pulse (section~\ref{sec:TOPS}).  If the dominant cause of energy jitter is intensity jitter, then we should gain additional suppression of $A_E$ from the IA loop.  Neither the half-wave plate nor the Asymmetry Inverter provides a cancellation of $A_E$.

\subsection{Collimation and Spot Size at CP Pockels Cell}
A converging or diverging laser beam traversing a Pockels cell acquires a gradient in its retardation that has an effect similar to a birefringence gradient as is discussed in section~\ref{sec:ss}.  Collimating the laser beam at the Pockels cells both minimizes the beam's angular spread and reduces the correlation between position and angle, suppressing these effects.  Increasing the beam diameter also reduces this effect but increases the sensitivity to birefringence gradient effects.  We compromise by placing a laser beam waist at the CP and PS cells with a size of $1\ \text{mm rms}$.

\subsection{Cathode Gradients}
Since we have observed that spatial variations in $\Delta_1$ and $\Delta_2$ lead to significant helicity-correlated position and spot size differences, it is natural to ask whether similar effects can arise from spatial variations in the other parameters of equation~\ref{eq:pita}, the analyzing power $\epsilon/T$ and the analyzer orientation $\theta$.  We speculate that the physical mechanism underlying the observation that position differences depend on choice of $\Delta_1$ and $\Delta_2$ along the $A_I=0$ line (section~\ref{strat_pita}) might be a manifestation of such variations.  If this is the case, then we can say that gradients in the cathode strain are a noticeable effect but clearly not the dominant effect in the SLAC system.  It is likely that suppressing $\Delta_1$ and $\Delta_2$ also suppresses $\alrb$'s arising from these variations.  However, neither the half-wave plate nor the Asymmetry Inverter provides a cancellation.

\subsection{Electronic Cross Talk}
Electronic cross talk between various necessary helicity-correlated signals and the DAQ can cause a mismeasurement of $\alrb$'s and the detector asymmetry.  Cross talk can be caused by coupling between either the transmission of the polarization state information or the high voltage pulse applied to the CP cell (which acts as an antenna) and ground loops in the DAQ.  The DAQ was carefully designed to avoid ground loops and we take the steps described in section~\ref{sec:DAQ} to eliminate the polarization state information as a problem.  Any residual effects are sensitive to cancellation by the half-wave plate.

\subsection{Etalon Effects}  Certain bench studies we conducted using an unwedged Pockels cell led us to consider the possibility that a Pockels cell with parallel faces can behave as a Fabry-Perot etalon.  A careful analysis of such a system reveals that it is capable of producing rather large intensity asymmetries (as large as $10^{-3}$ for a Pockels cell without AR coatings) on the laser beam, provided that the line width of the laser beam is small compared to the free spectral range of the Pockels cell etalon ($\sim3\ \text{GHz}$).  This, however, proves not to be the case for the SLAC system, for which the Flash:Ti laser has a line width of $\sim300\ \text{GHz}$.

\begin{longtable}{|>{\noindent}m{2.05in}|c|c|c|c|} 
\caption{Sources of $\alrb$'s.  The 2nd column indicates which $\alrb$'s can be affected (intensity asymmetry I, position (and angle) difference P, spot size difference S, or energy asymmetry E) and the 3rd column indicates the relative importance of this effect for the SLAC source. The 4th and 5th columns indicate if some cancellation of the effect is achieved by toggling the half-wave plate or the Asymmetry Inverter optics.\label{tab:table}} \\
\hline
\hline
Effect & $\alrb$'s & Significance & $\lambda$/2 Cancel & AI Cancel \\
\hline
\hline
Residual Linear Polarization coupling to QE Anisotropy\footnote{The coupling between residual linear polarization in the laser beam and the cathode QE anisotropy results in the dominant contribution to $A_I$.  It is also a dominant source of position and spot size differences and of the energy asymmetry via spatial gradients and beam loading (listed separately).} & I & dominant & NO & NO \\
\hline
 Birefringence Gradients & PS & dominant & Complex & position YES \\
\hline
 Lensing & PS & large & YES & position YES \\
\hline
 Beam Loading & E & large & NO & NO \\
\hline
 Collimation and Spot Size & IPS & medium & Complex & position YES \\
\hline
 Cathode Gradients & PS & medium & NO & NO \\
\hline
 Electronic Cross Talk & IPSE & small & YES & NO \\
\hline
 Etalon & I & negligible & NO & NO \\
\hline
\hline
\end{longtable}

%% file: n7.tex
\section{$\mathbf{Conclusion}$}
\setcounter{footnote}{0}
We have described the laser and optics systems for SLAC's polarized electron source for the $\text{E-158}$ parity violation experiment. This experiment seeks to measure a parity-violating right-left asymmetry with a statistical error of 8 parts per billion (ppb), and therefore requires systematic errors due to right-left beam asymmetries at the ppb level.  The laser is a stable, high-power, high-repetition rate, low-maintenance flashlamp-pumped Ti:Sapphire. Polarized electrons of either helicity can be generated by flipping the helicity of the laser beam. The optics system has been extensively analyzed and optimized to minimize $\alrb$'s.  The techniques we employ to achieve this include careful selection and setup of the polarization Pockels cells, imaging of the CP Pockels cell onto the cathode, active feedbacks on the helicity-correlated intensity asymmetry and position differences, half-wave plate reversals of the beam helicity, and an Asymmetry Inverter to reverse position and angle differences.  This system has been implemented for $\text{E-158's}$ physics run in 2002; it will be used for an additional $\text{E-158}$ physics run and for a variety of polarized beam experiments thereafter.

\section*{$\mathbf{Acknowledgements}$}
This work was supported by Department of Energy contracts DE-AC03-76SF00515 (SLAC), DE-FG02-01ER41168 (University of Virginia), DE-FG02-90ER40557 (Princeton University), DE-FG02-88ER40415 (University of Massachusetts), and DE-FG02-84ER4016 (Syracuse University), and by National Science Foundation Grant No. PHY-72192 (California Institute of Technology).